\documentclass[a4paper,11pt]{article}
\usepackage{jheppub} 
\usepackage{lineno}
\usepackage{amsmath}
\usepackage{amssymb}
\usepackage{physics}
\usepackage{xcolor}
\usepackage{subcaption}

\newcommand{\e}{\epsilon}

\newcommand{\even}{\text{even}}
\newcommand{\odd}{\text{odd}}

\def\be{\begin{equation}}
\def\ee{\end{equation}}
\def\bea{\begin{eqnarray}}
\def\eea{\end{eqnarray}}
\def\diag{\text{diag~}}
\def\subK{\text{{\tiny K}} }

\title{\boldmath Temperature dependence in Krylov space}

\author[a]{Nikolaos Angelinos} \author[b]{Debarghya Chakraborty} \author[c]{Anatoly Dymarsky}

\affiliation[a]{Yau Mathematical Sciences Center, Tsinghua University, Beijing 100084, China}

\affiliation[b]{ICTP South American Institute for Fundamental Research, Instituto de F\'isica Te\'orica UNESP
Rua Dr.~Bento Teobaldo Ferraz 271, 01140-070, S\~ao Paulo, SP, Brazil}

\affiliation[c]{Department of Physics, University of Kentucky, 506 Library Drive, Lexington, KY, 40506}

\abstract{
We consider the recursion method applied to a generic 2pt function of a quantum system and show, in full generality, that the temperature dependence of the corresponding Lanczos coefficients is governed by integrable dynamics. After an appropriate change of variables, Lanczos coefficients with even and odd indices are described by two independent Toda chains, related at the level of the initial conditions. Consistency of the resulting equations can be used to show that certain scale-invariant models necessarily have a degenerate spectrum. We dub this self-consistency-based approach the ``Krylov bootstrap''.
The known analytic behavior of the Toda chain at late times translates into analytic control over the 2pt function and Krylov complexity at very low temperatures.
We also discuss the behavior of Lanczos coefficients when the temperature is low but not much smaller than the spectral gap, and elucidate the origin of the staggering behavior of Lanczos coefficients in this regime.

}

\begin{document}
\maketitle
\flushbottom

\section{Introduction}
\label{sec:intro}
The Krylov space method, and Krylov complexity in particular, have emerged recently as a new way to explore and characterize quantum chaotic dynamics \cite{NANDY20251,Rabinovici:2025otw}. The Krylov space is defined as the minimal vector space that contains an operator $A$, together with its time evolution $A(t)$. It  is spanned by nested commutators of the operator with the Hamiltonian $[H,[H,\cdots [H,A]]]$. Utilizing the recursion method, the two-point function of $A$, that we will denote $C(t)$, can be rewritten in terms of  the so-called Lanczos coefficients $\{b_n\}$, that fully encode and characterize it.  Although the content of $C(t)$ and $b_n$ are mathematically equivalent, the behavior of $b_n$ was shown to  offer various insights into system's dynamics \cite{Parker19}. Thus, Krylov complexity $K(t)$, which is essentially 
the mean location of the operator $A(t)$ as it spreads along the ``Krylov chain'' -- a certain preferred basis in Krylov space -- has been proposed to characterize complexity growth and is conjectured to  
bound the OTOC growth \cite{Parker19}, generalizing the Maldacena-Shenker-Stanford bound on chaos \cite{avdoshkin2022krylov}.



Considering a system at  finite temperature $T=\beta^{-1}$,  Lanczos coefficients will exhibit temperature dependence $b_n(\beta)$ that encodes temperature dependence of the autocorrelation function $C_{\beta}(t)$. As one of the central results of this paper we find that $\beta$ dependence of $b_n$ is described by a completely integrable dynamical system, namely two uncoupled Toda chains, related at the level of initial conditions. This result is conceptually and technically similar to \cite{todaKrylov}, where the dependence of Lanczos coefficients  on Euclidean time was found to be governed by the Toda lattice equations.

An explicit form of the differential equations governing $\beta$-dependence of $b_n$ implies  various consistency conditions, giving rise to ``Krylov bootstrap''. These differential equations can also be used for numerical integration, calculating the thermal two-point function $C_\beta(t)$ starting from the  data associated with any other inverse temperature $\beta_0$.
Furthermore, recasting $\beta$-dynamics in terms of a completely integrable dynamical system provides 
analytic control in the large-$\beta$ limit, $\beta m \gg 1$, where $m$ is the spectral gap of the system. In this regime we find the asymptotic value of  Krylov complexity at late times to behave as $K\sim e^{-\beta m/2}$.

We also discuss the regime where $\beta$ is large but not much larger than $m^{-1}$. The integrable hierarchy does not yield fully universal predictions here, but we employ a number of exact mathematical results to explain the origin of the so-called staggering behavior, where $b_n$ splits into two smooth branches, observed empirically earlier in \cite{Mitra,Mitra2,Mitra3,Yeh:2023fek,avdoshkin2022krylov,Camargo_2023}. 

The paper is organized as follows. In \ref{section2} we derive the analytic equations driving the $\beta$-dependence of $b_n$. Section \ref{sec:app} makes use of these results by studying the large temperature limit, as well as evaluating $C_\beta(t)$ numerically starting from $\beta=0$ data. Section \ref{sec:moderate} is devoted to the moderately large-$\beta$ regime, where we use simple analytic models to describe characteristic behavior of $b_n$ exhibited by many spatially-extended systems. We conclude with a discussion in section \ref{sec:discussion}. 


\section{Temperature dependence of Lanczos coefficients as a completely  integrable system}\label{section2}
The starting point is the Lanczos algorithm, which is used to iteratively build an orthogonal basis in the Krylov space, starting with an initial operator $A$. Orthogonality is defined with respect to a choice of an inner product, defined in terms of  the correlation function. Our goal is to understand temperature dependence of the correlation function and we define the scalar product \be \langle A,B\rangle_{\beta} \equiv \Tr(A^\dagger \rho_1(\beta) B\rho_2(\beta) ). \label{definip} \ee
such that the autocorrelation function $C_\beta(t)$ and spectral function $\Phi(\omega)$ are given by
\be
C_{\beta}(t) = \langle A(0), A(t) \rangle_{\beta} =\int_{-\infty}^\infty d\omega~ \Phi(\omega)e^{i\omega t}.\label{ipandauto}
\ee
In general $\rho_1(\beta)$ and $\rho_2(\beta)$ are Hermitian positive semi-definite operators that commute with the system's Hamiltonian. In most of this work we  focus on the Wightman inner product by making the choice
\be \rho_1=\rho_2={e^{-{\beta\over 2}H}}, \label{wip}\ee
where $H$ is the Hamiltonian of the quantum  system. 
Later in \ref{degeneracysection}, we will explain how our formalism can be straightforwardly adapted to the choice of other thermal inner products.  For an initial Hermitian operator $A_0=A$, we define the Krylov basis recursively
\be A_{n+1}=[H,A_n]-b_{n-1}^2 A_{n-1},\label{lanczos}\ee
where we set $b_{-1}= 0$.
The Lanczos coefficients $b_n(\beta)$ are fixed by requiring that the basis is orthogonal,
\be \label{b2def}
b_n^2(\beta)={\langle A_{n+1},A_{n+1}\rangle_{\beta} \over \langle A_n,A_n\rangle_{\beta}}.\ee
Since the norms are non-negative, for convenience, we define $q_n(\beta)$ as follows
\be 
\delta_{nm}e^{q_n(\beta)}:=\langle A_n,A_m\rangle_{\beta} \label{defqn}.\ee
It is sometimes convenient to work with the orthonormal Krylov basis $\{O_n\}$ defined by
\be O_n(\beta):={A_n\over \sqrt{\langle A_n,A_n\rangle}_{\beta}}={A_n}(\beta) e^{-{q_n(\beta) \over2}}.\ee
Since the inner product (\ref{wip}) is $\beta$-dependent, all quantities introduced above, including the Lanczos coefficients $b_n(\beta)$ and the Krylov basis operators $O_n(\beta)$, are $\beta$-dependent. 
Our main goal is to understand this temperature dependence. We will often suppress the explicit $\beta$-dependence in our notation, except where necessary for clarity.

From (\ref{lanczos}) it follows that the action of the Liouvillian $[H,\cdot]$ on the normalized Krylov basis has a  tridiagonal representation, 
\be M_{nm}(\beta):=\langle O_n(\beta),[H,O_m(\beta)]\rangle_{\beta}=\begin{pmatrix}
    0 & b_0 & 0& 0 & \ddots\\
    b_0 & 0 & b_1& 0 & \ddots\\
        0 & b_1 & 0 & b_2 & \ddots\\
    0 & 0 & b_2 & 0 & \ddots\\
    \ddots & \ddots & \ddots & \ddots & \ddots
\end{pmatrix}. \label{Liouvillian}\ee
Central to our work is the point that regardless of the temperature, which defines the scalar product in the Krylov space, the Krylov space itself, and the adjoint action of $H$ are temperature-independent. This implies then that $M(\beta)$ for different values of $\beta$  are different representations of the same operator, thus $\beta$-dependence is an isospectral deformation of $M$ which we will describe in terms of integrable dynamics. 

To this end  it is convenient  to introduce ``temperature evolution'' by considering $\beta$-dependent operators,
\be
D[\beta] := e^{-\frac{\beta}{2} H }\, D\, e^{-\frac{\beta}{2} H },   \label{tempdynamics}
\ee 
for any operator $D$. 
An important subtlety is that temperature evolution of $A=A_0$ under \eqref{tempdynamics} will in most cases move it outside of the Krylov space. To see that, we decompose $A$ as follows,
\be
A = A^{\perp} +  A_{\rm d}, \label{decomp} 
\ee 
where $A_{\rm d}$ and $A^{\perp}$ stand for  diagonal and off-diagonal parts of $A$, written in the eigenbasis of $H$. 
The diagonal operator $A_{\rm d}$ is in the nullspace of the Liouvillian and therefore all operators $A_n$ defined by \eqref{lanczos} with the same parity of $n$
will share the same diagonal part (up to normalization). On the contrary, the diagonal part of $A[\beta]$ is $\beta$-dependent and hence for $\beta\neq 0$ is outside the Krylov space. 
This problem appears for any degenerate energy gap $\omega_{ij}=E_i-E_j$; it is always present for zero energy gap $\omega=0$ because the latter is always maximally degenerate. Keeping this in mind, in full generality we can introduce the  superoperator $-{1\over 2}\{H , . \}$ acting as follows 
\be 
-{1\over 2}\{H , O_n \} =:\sum_m \mathcal{T}_{nm} O_m + \mathcal{T}_n^{\perp}  \implies \mathcal{T}_{nm}(\beta) := -\frac{1}{2} \langle O_m(\beta) , \{H , O_n(\beta) \} \rangle_{\beta} ,\label{Jrepgen} 
\ee 
where the matrix $\mathcal{T}$ represents the action of the above superoperator within the Krylov space and by $\mathcal{T}_n^{\perp}$ we denote the components not contained in the Krylov space. 

Our goal will be to derive differential equations governing the $\beta$-dependence of $\mathcal{T}_{nm}$ and $M_{mn}$.
We will first assume that $A$ has no non-zero matrix elements $A_{ij}$ corresponding to degenerate gaps $\omega_{ij}$, in particular, all diagonal matrix elements of $A$ have to be zero. This ensures that all $\mathcal{T}_n^{\perp}$ vanish. This scenario would normally apply to a chaotic $H$ with no non-zero gaps that are degenerate, and $A$ with vanishing thermal expectation value at all temperatures, due to some discrete symmetry. We then generalize to include  arbitrary Hermitian $H$ and $A$. 

\subsection{No degenerate energy gaps}
\label{subsection1}
First we consider the case when $\{H,\cdot\}$  does not move operators outside the Krylov space, which means all $\mathcal{T}_n^{\perp}$ vanish. Anticipating the need to track components of the superoperator $-\frac{1}{2} \{ H, . \}$ within the Krylov space separately in the more general case, we introduce the matrix $T$ which accomplishes this. In the present case, $T=\mathcal{T}$.  It is easy to see that the matrices $T(\beta),M(\beta)$ commute, as follows from 
\be [H,\{H, \,.\, \}]=\{H,[H, \,.\,  ]\}.\ee
Next, we temperature-evolve the basis operators of the Krylov space
\be 
O_n(\beta)\left[{\beta-\beta_0\over 2}\right]=
e^{-{(\beta-\beta_0) H\over 4}}O_n(\beta)e^{-{(\beta-\beta_0) H\over 4}}=\sum_m \left(e^{T(\beta){\beta-\beta_0\over 2}}\right)_{nm}O_m(\beta) \label{TevolvedKryl}. \ee
It is clear that $O_n(\beta)\left[{\beta-\beta_0\over 2}\right]$ are mutually orthogonal with respect to the scalar product $\langle\cdot , \cdot \rangle_{\beta_0}$. Thus, the basis $\{O_n(\beta)\left[{\beta-\beta_0\over 2}\right]\}$ is related to the basis $\{O_n(\beta_0)\}$ by an orthogonal transformation $Q$, 
\be \sum_{m}(e^{T(\beta){\beta-\beta_0\over 2}})_{nm}O_m(\beta)=\sum_{m} Q_{nm}(\beta,\beta_0)O_m(\beta_0). \label{basischange}\ee
Acting on both sides of (\ref{basischange}) with $[H,\cdot]$, and using $[M,T]=0$, we obtain 
\be M(\beta)=Q(\beta,\beta_0) M(\beta_0)Q^T(\beta,\beta_0).\label{isoM}\ee
Similarly, acting by $-{1\over 2}\{H,\cdot\}$ and using (\ref{Jrepgen}) we obtain
\be T(\beta)=Q(\beta,\beta_0) T(\beta_0)Q^T(\beta,\beta_0).\label{isoT}\ee
As was expected, $\beta$-evolution of both $M$ and $T$ is an isospectral deformation and can be written in Lax form. 
The ``evolution'' operator $Q$ can be written as $Q(\beta,\beta_0)=U(\beta)U^\dagger(\beta_0)$, where $U$ is an orthogonal matrix. 
The recursion relation \eqref{lanczos} implies that the basis element $O_n(\beta)$ is a linear combination of the first $n$ elements of the basis $O_m(\beta_0)$. Hence there is a lower-triangular matrix $R^T$ such that
\be
O_n(\beta)=\sum_m (R^{-1})^T_{nm}(\beta,\beta_0)O_m(\beta_0)
\ee
Therefore, (\ref{basischange}) can be written as follows
\be \sum_{m}(e^{T(\beta){\beta-\beta_0\over 2}})_{nm} O_m(\beta)=\sum_{m,k} Q_{nm}(\beta,\beta_0)R^T_{mk}(\beta,\beta_0)O_k(\beta)\implies e^{T(\beta){\beta-\beta_0\over 2}}=Q(\beta,\beta_0)R^T(\beta,\beta_0), \label{basischange2}\ee
where $R$ is an upper triangular matrix. Using (\ref{isoT}) with (\ref{basischange2}) we find the QR decomposition
\be e^{T(\beta_0){\beta-\beta_0\over 2}}=Q^T(\beta,\beta_0) R(\beta,\beta_0).\label{QR}\ee 
Taking a derivative with respect to $\beta$ and using (\ref{isoT}), we obtain
\be {1\over 2} T(\beta)=-B(\beta) +\dot R(\beta,\beta_0) R^{-1}(\beta,\beta_0),\label{TR}\ee
where we defined
\be B(\beta)=-Q(\beta,\beta_0)\dot Q^T(\beta,\beta_0).\ee
Notice, that $B$ is independent of $\beta_0$.
Now using the fact that $B$ is antisymmetric and $\dot R R^{-1}$ is upper triangular, (\ref{TR}) implies that
\be B(\beta)={1\over 2}(T^+(\beta)-T^-(\beta))  \label{Bdef},\ee
where by $T^+,T^-$ we denote the upper-triangular and lower-triangular parts of $T$, respectively. Finally, taking a derivative of (\ref{isoM}) and (\ref{isoT})  with respect to $\beta$ gives
\be 
{dT\over d\beta}\equiv \dot T=[B,T], \quad  {dM\over d\beta}\equiv\dot{M}=[B,M],\quad B={1\over 2}(T^+-T^-). \label{TandMdot}\ee
The equations (\ref{TandMdot})  describe completely integrable dynamics. They govern the temperature-dependence of Lanczos coefficients $b_n(\beta)$. This is one of our main results, written in a simplified scenario of no degenerate energy gaps. We note, that similar equations  would describe the dependence of $b_n$ on any continuous parameter deforming the scalar product. 

Since our main goal is to understand temperature dependence of $b_n$, or equivalently, $q_{n}(\beta)$, we would like to parameterize $T$ with the same variables. From (\ref{Jrepgen}), and using the fact than the unnormalized bases $A_n(\beta)$ and $A_n(\beta_0)$ are related by a lower-triangular matrix with $1$ in its diagonal, it follows that the diagonal entries of $T$ are given by
\be h_n:= T_{nn} = \dot{q}_{n}. \ee 
One possible strategy is to make use of $\comm{M}{T} = 0$ to recursively solve for the entries of $T$ in terms of $h_n$ and the Lanczos coefficients $b_n$. We can build $T_{nm}$ recursively, starting from the diagonal, as follows
\be T_{n, n+(k+2)} =  \frac{b_{n-1} T_{n-1, n-1 + (k+2)} + b_n T_{n+1, n+1 + k} - b_{n+k} T_{n, n+k} }{b_{n+1+k}}. \label{Trec} \ee

Care must be taken to ensure the proper termination condition in the case of finite-dimensional Krylov space of size $N$, namely in equation \eqref{Trec} it is understood that $b_n=0$ for $n\geq N$ and $n<0$. The Lax equations (\ref{TandMdot}) can be written as a system of differential equations for $b_n,h_n$, i.e.~for both $q_n(\beta)$ and $\dot{q}_n(\beta)$. To specify a solution, one needs to supply initial conditions not only for $b_n$, but also for $h_n$. The former can be evaluated using  Lanczos algorithm at the initial value $\beta=\beta_0$. To evaluate $h_n$ one needs to evaluate the following quantities in the initial Krylov basis at $\beta=\beta_0$,
\be h_n(\beta_0)=-{1\over 2}\langle O_n(\beta_0),\{H,O_n(\beta_0)\}\rangle_{\beta_0}.\ee
An explicit example of writing down and solving differential equations for $q_n(\beta)$  is given in subsection \ref{subsectionexample}.

Finally we can identify the orthogonal matrix $U(\beta)$ introduced above with the matrix that diagonalizes $M$ 
\be M(\beta)=U(\beta)\Lambda U^T(\beta ). \label{defU} \ee
Here  $\Lambda$ is a $\beta$-independent diagonal matrix of eigenvalues, and $U$ satisfies

\be U(\beta)= Q(\beta,\beta_0)U(\beta_0),\qquad \dot U(\beta)=B(\beta) U(\beta).\ee

\subsubsection{Relation to Toda chain}\label{sec:Toda}
There is another way to build a basis in the Krylov space. We start by performing the Lanczos algorithm with the super-operator $-{1\over 2}\{H,\cdot\}$ and the initial operator $\tilde A_0^\text{even}\equiv A_0$:
\be \tilde A^\text{even}_{n+1}=-{1\over 2}\{H,\tilde A^\text{even}_n\}-\tilde \alpha^\text{even}_n \tilde A^\text{even}_n-(\tilde b^\text{even}_{n-1})^2 \tilde A_{n-1}^\text{even},\label{Lak}\ee
where
\be (\tilde b_{n}^\text{even})^2={\langle \tilde A^\text{even}_{n+1},\tilde A^\text{even}_{n+1}\rangle_\beta\over \langle \tilde A^\text{even}_n,\tilde A^\text{even}_n\rangle_\beta},~\tilde \alpha^\text{even}_n={\langle \tilde A^\text{even}_n,\{H,\tilde A^\text{even}_n\}\rangle_\beta\over\langle \tilde A^\text{even}_n,\tilde A^\text{even}_n\rangle_\beta}.\label{Lak2}\ee
After normalizing $\tilde{A}^\text{even}_{n}$ we obtain $\tilde O^\text{even}_n$. We observe that for a Hermitian initial operator $A_0$, the Krylov basis $\{ O_n \}$ for even and odd $n$  splits into Hermitian and anti-Hermitian operators, respectively. However, $\{H,\cdot\}$ maps a Hermitian (anti-Hermitian) operator to another Hermitian (anti-Hermitian) operator. This implies that $\{\tilde O^\text{even}_n\}$ spans only half of the Krylov space, namely the subspace spanned by Hermitian operators. To obtain the rest of the Krylov space, we must perform the Lanczos procedure again, but this time starting with the initial operator $\tilde A^\text{odd}_0\equiv A_1$. This leads to another set of operators $\{\tilde O^\text{odd}_n\}$ and Lanczos coefficients $\tilde b^\text{odd}_n,\tilde a^\text{odd}_n$.

The super-operator $-{1\over 2}\{H,\cdot\}$ written in the bases $\{\tilde O^\text{even}_n\}$, $\{\tilde O^\text{odd}_n\}$  has tridiagonal representations $\tilde T_\text{even},\tilde T_\text{odd}$, 
\be (\tilde T_\text{even})_{nm}=-{1\over 2}\langle \tilde O^\text{even}_n,\{H,\tilde O^\text{even}_m\}\rangle ,\ee
\be (\tilde T_\text{odd})_{nm}=-{1\over 2}\langle \tilde O^\text{odd}_n,\{H,\tilde O^\text{odd}_m\}\rangle ,\ee
while $\langle \tilde O^\text{odd}_n,\{H,\tilde O^\text{even}_m\}\rangle=\langle \tilde O^\text{even}_n,\{H,\tilde O^\text{odd}_m\}\rangle=0$.
Following the same steps as in \cite{todaKrylov}, it is straightforward to show that the temperature evolution of these matrices is governed by Toda equations
\be \dot {\tilde{T_i}}=[\tilde B_i,\tilde T_i],~\tilde B_i={1\over 2}(\tilde T_i^+-\tilde T_i^-)~~~ i=\text{even, odd}.\label{Toda}\ee
Note that we use the index $i$ to label the $\tilde{T}$ in different spaces and indices $m,n$ to denote matrix elements.  

The elements of the Krylov basis $\{ O_n \}$  alternate between Hermitian and anti-Hermitian operators for even and odd $n$ respectively so that $T_{nm}=0$ unless $n+m$ is even. Hence, similar to the case of $\tilde{T}$, we can split $T$ into a direct sum of even and odd parts
\be (T_\even)_{nm}=T_{2n,2m} ~~ (T_\odd)_{nm}=T_{2n+1,2m+1}.\ee
We rewrite equation \eqref{TandMdot} as two Lax equations
\be \dot T_i=[B_i,T_i],~~B_i={1\over 2}(T_i^+-T_i^-),~~i=\even,\odd.\label{Tdotevenodd}\ee
Therefore the dynamical equations governing even and odd subspaces are fully decoupled. It follows from the Lax representation that the constants of motion are 
\be \mathcal I_k^{i}={1\over k} \tr(T_i^k) = {1\over k} \tr(\tilde{T}_i^k) ,~~~i=\even,\odd \label{constofmotion} \ee
The condition that $T$ and $M$ commute, in terms of $T_\text{even},T_\text{odd}$  can be written as follows 
\be T_\text{even}\,{\cal M}={\cal M}\,T_\text{odd},\quad {\cal M}=\begin{pmatrix}
    b_0 & b_1 & 0 &\cdots\\
    0 & b_2 & b_3&\cdots\\
    0 & 0 & b_4 &\cdots\\
    \vdots&\vdots &\vdots &\ddots
\end{pmatrix}.\label{mdef}\ee
This implies that ``on-shell,'' i.e.~for the initial conditions specified by  \eqref{constofmotion}, 
\be \mathcal I_k^\even=\mathcal I_k^\odd. \ee
Equations (\ref{Tdotevenodd}) and  \eqref{Toda} have the same functional form. 
They are in fact equivalent. To see that we note that 
matrices $T_\even,T_\odd$ for any given $\beta$ 
are real-symmetric and can be tri-diagonalized using Lanczos algorithm with the $\beta$-independent initial vector $v_0=(1,0,\dots,0)$ representing operators $A_{0}, A_1$  for $i={\rm even}$ and $i={\rm odd}$ respectively. 
In other words $\tilde T_l(\beta)$, for $l=\even,\odd$ is  $T_l(\beta)$ brought to the tridiagonal form by the orthogonal transformations $P_l(\beta)$, 
\be 
\tilde{T}_l(\beta) =P_l(\beta) T_l(\beta) P^{T}_{l}(\beta),~~~l=\even,\odd.\label{Pevenodd}
\ee
Orthogonal matrix $P_l$ is uniquely fixed by the requirement that $\tilde{T}_l$ is tri-diagonal and $P_l$ is block-diagonal such that $(P_l)_{00}=1$ and $(P_l)_{0n}=(P_l)_{n0}=0$ for any $n\neq 0$. Comparing (\ref{Tdotevenodd}) and  \eqref{Toda} we find that
\bea
\tilde{B}=P_l(\beta) B_l(\beta) P^{T}_{l}(\beta)+\dot{P}_l P_l^T,
\eea
and this equation is consistent with $B_l,\tilde{B}_l$ being related to $T_l,\tilde{T}_l$ as given by (\ref{Tdotevenodd}) and \eqref{Toda}. In other words, starting with some $T_l(\beta_0)$ one can first evolve it in $\beta$ using $\dot{T}_l=[B_l,T_l]$ to $\beta_1$ and then tridiagonalize using Lanczos algorithm with $v_0$ as a starting vector, or first tridiagonalize using the same initial vector and then evolve using  $\dot{\tilde{T}}_l=[\tilde{B}_l,\tilde{T}_l]$ from $\beta_0$ to $\beta_1$. In both cases the result will be the same, see \cite{deift1986toda,bloch2023symmetric} for a formal proof.  

Going back to \eqref{Toda}, these equations describe two decoupled open Toda chains, related only by the initial conditions $\mathcal I_k^\even=\mathcal I_k^\odd$ for all $k$.  In terms of the phase-space variables $\{\tilde{p}^{l}=\dot{\tilde{q}}^l, \tilde{q}^{l}   \}$ with the canonical Poisson bracket, the Hamiltonian generating \eqref{Toda} is
\be {\mathcal H}={\mathcal I}_2^\even+{\mathcal I}_2^\odd=\tilde{\mathcal I}_2^\even+\tilde{\mathcal I}_2^\odd. \label{TodaH} \ee
The same Hamiltonian \eqref{TodaH} generates the dynamics of the original matrix $T$. However, the relation between $q_n(\beta)$ and $\tilde{q}_n(\beta)$ is non-universal and complicated. Both the Hamiltonians  ${\mathcal I}_k^{l}$ and the Poisson brackets look non-trivial in terms of $\dot{q}_n, q_n$.

\subsubsection{Example I}\label{subsectionexample}

As a simple example, let us consider two coupled spins governed by the Hamiltonian
\be H=-J(\sigma_x^1\sigma_y^2+\sigma_y^1\sigma_x^2)+g(\sigma_y^2+\sigma_x^1).\ee
The eigenvalues of this Hamiltonian are
\be E_0=0,~E_1=2J,~E_2=-J-\sqrt{J^2+4g^2},~E_3=-J+\sqrt{J^2+4g^2}.\ee
Starting from the initial operator, which has zero diagonal elements in the energy eigenbasis,
\be A_0=\sigma_y^1\sigma_z^2,\ee
and performing  Lanczos algorithm with $\beta=0$ we find
\begin{eqnarray} 
A_1&=&4ig(\sigma_y^1\sigma_x^2+\sigma_z^1\sigma_z^2)+4i J\sigma_y^2,\\
A_2&=&-16g^2\sigma_z^1\sigma_x^2,\\
A_3&=&32i{g^2J\over 2g^2+J^2}\left((2g^2+J^2)\sigma_x^1+gJ\sigma_y^1\sigma_x^2-2g^2\sigma_y^2+gJ\sigma_z^2\sigma_z^2\right),
\end{eqnarray}
along with Lanczos coefficients
\be b_0(0)=2\sqrt{J^2+2g^2},\quad b_1(0)={4g^2\over \sqrt{2g^2+J^2}},\quad b_2(0)=2 \sqrt{\frac{4 g^2 J^2+J^4}{2 g^2+J^2}}.\label{betas}\ee
We also evaluate
\be h_i(0)=-{1\over 2}\tr(A_i^\dagger \{H,A_i\})/\tr(A_i^\dagger A_i),\ee
resulting in
\be h_0(0)=h_2(0)=0,~h_1(0)=-h_3(0)={2g^2 J\over 2 g^2 + J^2} .\label{hs}\ee

The Krylov space is 4-dimensional. Let us write
\begin{eqnarray}  
&&T^\text{even}=\begin{pmatrix}
	h_0 & t_{02}\\
	t_{02} & h_2
\end{pmatrix},\qquad T^\text{odd}=\begin{pmatrix}
	h_1 & t_{13}\\
	t_{13} & h_3
\end{pmatrix},\qquad M=\begin{pmatrix}
    0 &b_0 & 0 &0\\
    b_0 & 0 & b_1 & 0\\
    0 &b_1 & 0 & b_2\\
    0 & 0 &b_2 & 0
\end{pmatrix}\\
&&t_{02}={b_0\over b_1}(h_1-h_0),\quad t_{13}={b_0^2(h_1-h_0)+b_1^2(h_2-h_1)\over b_1b_2},\end{eqnarray}
where we used (\ref{Trec}) to determine the off-diagonal entries of matrix $T$ in terms of $b_i,h_i$.

The Lax equations $\dot T=[B,T]$, $\dot M=[B,M]$ become the following system of differential equations
\be \dot h_0=-\dot h_2={b_0^2\over b_1^2}(h_0-h_1)^2,~~\dot h_1=-\dot h_3={b_2^2\over b_1^2}(h_2-h_3)^2,\ee
\be \dot b_0={1\over 2}b_0(h_1-h_0),~\dot b_1={1\over 2}b_1(h_2-h_1),~\dot b_2={1\over 2}b_2(h_3-h_2).\ee
Taking advantage of the fact that the quantities $\tr(M^{2})$ and $\tr((T^\text{even})^k)=\tr((T^\text{odd})^k)$ for $k=1,2$ are conserved, we find the general solution
parameterized by the constants $c,\kappa, v_0,v_1,\lambda$,
\be h_0={c\over 2}+\kappa \tanh(\kappa(\beta-v_0)),\ee
\be h_2={c\over 2}-\kappa \tanh(\kappa(\beta-v_0)),\ee
\be h_1={c\over 2}+\kappa \tanh(\kappa(\beta-v_1)),\ee
\be h_3={c\over 2}-\kappa \tanh(\kappa(\beta-v_1)),\ee
\be b_0={\lambda \over \sqrt{ \cosh \left(\kappa(\beta+v_1-2v_0) \right) \text{sech}\left(\kappa(\beta
		-v_1) \right)+1}},\ee
\be b_1={\lambda~\text{sech}(\kappa(\beta-v_1))|\text{sinh}(\kappa(v_1-v_0))| \over \sqrt{ \cosh \left(\kappa(\beta +v_1-2v_0) \right) \text{sech}\left(\kappa(\beta
		-v_1) \right)+1}} ,\ee
\be b_2={\lambda \cosh(\kappa(\beta-v_0))\text{sech}(\kappa(\beta-v_1)) \over \sqrt{ \cosh \left(\kappa(\beta +v_1-2v_0) \right) \text{sech}\left(\kappa(\beta
		-v_1) \right)+1}} .\ee

Given the initial values  (\ref{betas},\ref{hs}), the integration constants can be evaluated to be

\be c=0,~\lambda=2\sqrt{2J^2+4g^2},~\kappa=-J,~v_0=0,~v_1=-{1\over J}\text{arctanh}{2g^2\over 2g^2+J^2} .\ee

This completes the derivation of $b_n(\beta)$.

Asymptotically, as $\beta\to\infty$ we find that $b_0\to |E_1-E_0|$, $b_1\to 0$, $b_2\to |E_3-E_2|$. This is a pattern that we explore in generality in section \ref{asymptoticsec}.

We note that in this simple example, $T_\text{even}=\tilde T_\text{even}$, $T_\text{odd}=\tilde T_\text{odd}$, so this system is exactly the same as two conventional Toda chains of size two, with the dynamical variables $\{h_i=\dot{q}_i,q_i\}$ and the canonical  Poisson brackets. This is no longer true for Krylov spaces of larger sizes, and the Poisson brackets  $\{q_i,h_j\}$ become complicated.

\subsubsection{Example II}
\label{exactsol}
In this section we consider another analytic example of $\beta$-dependent $b_n$, when the Krylov space is infinite-dimensional. The shortcoming of this example is that, unlike the example considered above, it is not associated with any physical system. 

As a starting point we assume that for all $\beta$, the matrix $T$ is functionally related to 
$M$,  $T=f(M)$, for some appropriate function $f$. Since $T$ is symmetric and $M$ is antisymmetric, function $f$ is even. A simple example is given by 
$f(M) = \alpha\, M^{2k}$ with fixed $\alpha,k$. 
Then the equation $\dot{M} = \comm{B}{M}$ becomes an instance of the Toda hierarchy. 

The case with $T = -\frac{1}{2} \alpha\,  M^2$ is particularly interesting. In this case  $T$ is a direct sum of two coupled Toda chains $T_{\odd}$ and $T_{\even}$ and the expression for $T(\beta)$ can be found explicitly. Using our notation  for $T_{\text{even}}$, we find $({b_{\even}})_{n} = b_{2n}b_{2n+1}$ and $(a_{\even})_n = b^2_{2n} + b_{2n-1}^2 $  and similarly for  $({b_{\odd}})_{n} = b_{2n+1} b_{2n+2}$ and $(a_{\odd})_n= b^2_{2n} + b_{2n+1}^2$. A simple calculation shows  that $\dot{T} = [B, T] $ reduces to two decoupled Toda equations for $T_{\even}$ and $T_{\odd}$. Writing it instead in terms of $b_n$, we find the integrable system studied by Kac and Van Moerbeke \cite{KacVanMoerbeke}

\be 
\dot{b}_{n}(\beta) = - \frac{\alpha}{2} b_n (\beta)( b^2_{n+1}(\beta) - b^2_{n-1}(\beta) ).      \label{KacVanMoerbeke}
\ee 
There is a  separable polynomial solution to \eqref{KacVanMoerbeke} given by
\be
b_n(\beta) = \alpha \sqrt{ \frac{n+1}{\beta - \beta_0} }.  \label{exactsolution}
\ee 
The``autocorrelation function'' associated with these Lanczos coefficients  is simply the Gaussian
\be
\label{cb}
C_{\beta}(t) = e^{-\frac{\alpha^2 t^2}{2 (\beta - \beta_0) } }. 
\ee 
While the time dependence of $C_{\beta}(t)$ is perfectly physical and is exhibited by many systems, we do not know any system that would have temperature dependence as in\eqref{cb}.

\subsection{General initial operator, no degenerate energy gaps} \label{sssec:general}
For a general Hermitian initial operator with non-vanishing diagonal matrix elements $A_{ii}$, the derivation of the preceding section will be altered, even if the energy spectrum has no non-trvial degenerate gaps. This is because the  zero gap is always degenerate. Therefore  the Krylov spaces obtained from $A[\beta]$ will depend on $\beta$. However, it is possible to keep track of the diagonal parts of the operators $A_n$ separately.  We modify the definition of $T$ to
\be 
T_{nm}(\beta)=-{1\over 2}\langle O_n,\{H,O_m\}\rangle_{\beta}-{1\over 2}a_m\langle O_n,\{H,e^{-\mu(\beta,\beta_0) }d(\beta_0)\}\rangle_{\beta},\label{Tnm}\ee
\be d(\beta_0) := \frac{\diag A}{~~ \sqrt{\langle \diag A, \diag A \rangle_{ \beta_0}}}, \quad e^{2\mu(\beta,\beta_0)}:=\langle d,d \rangle_{\beta}, \label{ddef} \ee 
where, up to normalization, $d$ is the diagonal part of $A$ -- the projection of $A$ to the nullspace of $[H,\cdot]$ (we assume that the energy spectrum is non-degenerate such that each energy gap $\omega_{ij}=E_i-E_j$, except for $\omega_{ii}=0$, is not degenerate). The function $\mu(\beta,\beta_0)$ is defined by the thermal two-point function of the diagonal part of the initial operator. In \eqref{Tnm} the basis operators $O_n$ are $\beta$-dependent. The vector $a_n$ is the normalized null-vector of the Liouvillian $M$, and consequently also the null-vector of $T$. Explicitly, $a_n$ can be written  in terms of   Lanczos coefficients as follows
\be 
a_n(\beta) =\begin{cases}
   (-1)^k \bigg( \prod^{l=k-1}_{l=0} \frac{b_{2l}(\beta)}{b_{2l+1}(\beta) }\bigg) a_0(\beta) ,  &n=2k ,\\
  0  ,    &n=2k+1, 
\end{cases}\label{ab}
\ee 
and $a_0(\beta)$ is given by 
\be a_0(\beta) = \frac{\langle d, O_0 \rangle_{\beta}}{\sqrt{\langle d, d \rangle_{\beta}}}.  \ee

With these modifications, $T$ is equal to  $-{1\over 2}\{H,\cdot\}$ projected on the Krylov space with the direction corresponding to the degenerate eigenvalue of $M$ subtracted, while the evolution of the initial operator outside this space is governed by the functions $a_n(\beta)$ and $\mu(\beta)$. Matrices $T$ defined in \eqref{Tnm} and  $\mathcal{T}$ introduced in \eqref{Jrepgen} are related by the shift 
\be T_{nm} = \mathcal{T}_{nm} -{1\over 2}a_m\langle O_n,\{H,e^{-\mu(\beta,\beta_0) }d(\beta_0)\}\rangle_{\beta} = \mathcal{T}_{nm}-2\dot{\mu} a_n a_{m}. \label{relationbetweenTs}  \ee 
This equation  follows from the observation that the inner product in the rightmost term in \eqref{Tnm} only gets a contribution from the diagonal part of $O_n \propto a_n d$. The proportionality constant can be written as a $\beta$ derivative of \eqref{ddef} giving rise to  $\dot{\mu}$ in \eqref{relationbetweenTs}. 

From (\ref{Tnm}) it follows that the temperature evolution operator acting on the Krylov basis can be expressed as
\be e^{-{(\beta-\beta_0) H\over 4}}O_n(\beta)e^{-{(\beta-\beta_0) H\over 4}}=\sum_m (e^{T(\beta){\beta-\beta_0\over 2} + \mu(\beta,\beta_0) a(\beta) a(\beta)^{T}  })_{nm}O_m(\beta) \label{TevolvedKryl2}. \ee
Similarly to (\ref{QR}), we consider the QR decomposition
\be e^{T(\beta_0){\beta-\beta_0\over 2} + \mu(\beta,\beta_0) a(\beta_0) a^T(\beta_0)  }=Q^T(\beta,\beta_0) R(\beta,\beta_0).\label{QR2}\ee
The matrix $T$ and its null-vector $a$ evolve as
\be T(\beta)=Q(\beta,\beta_0) T(\beta_0)Q^T(\beta,\beta_0),~~a(\beta)=Q(\beta,\beta_0)a(\beta_0).\label{isoTa}\ee

Taking the derivative of (\ref{QR2}) with respect to $\beta$ we find that the Lax equations \eqref{TandMdot} continue to hold with a generalized $B$: 
\bea B &=&  {1\over 2}(T^+-T^-)+\dot\mu(\beta)((aa^T)^+-(aa^T)^-),  \label{generalB} \\
\dot{M}&=&[B,M],\qquad \dot{T}=[B,T]. \label{MTdynamics}
\eea 
We note that, due to the definition (\ref{ddef}), $\dot\mu(\beta)\equiv{\partial\over \partial\beta}\mu(\beta,\beta_0)$ is independent of $\beta_0$.
Just like before, the equations for the odd and even subspaces  decouple
\begin{eqnarray}
\dot{T}_i&=&[ B_i, T_i], ~~~i=\even,\odd, \label{lax2}\\ \nonumber
B_\even&=&{1\over 2}(T_\even^+-T_\even^-)+\dot\mu((aa^T)_\even^+-(aa^T)_\even^-), \quad B_\odd={1\over 2}(T_\odd^+-T_\odd^-).\end{eqnarray}
These equations do not determine the function $\dot\mu(\beta)$, which should be evaluated independently from its definition (\ref{ddef}).

The differential equations above can be recast as Hamiltonian dynamics. To see this, we can adapt the analysis of section \ref{sec:Toda}. After building the bases, $\{\tilde O_n^\even\}$ and $\{\tilde O_m^\odd\}$, we define the triadiagonal matrices 
\begin{eqnarray}
(\tilde T_\even)_{nm}(\beta)&=&-{1\over 2}\langle \tilde O^\even_n,\{H,\tilde O^\even_m\}\rangle_{\beta}-{1\over 2}\tilde a_m\langle \tilde O^\even_n,\{H,e^{-\mu(\beta,\beta_0) }d(\beta_0)\}\rangle_{\beta},\\
(\tilde T_\odd)_{nm}(\beta)&=&-{1\over 2}\langle {\tilde O^\odd}_n,\{H,{\tilde O^\odd}_m\}\rangle_{\beta},
\end{eqnarray} 
where $\mu(\beta,\beta_0)$ is defined in \eqref{ddef} and $\tilde a$ is the normalized null-vector of $\tilde T_{\text{even}}$. Writing the equations for even and odd subspaces analogous to \eqref{TevolvedKryl2} and following the same procedure, we find that $\tilde T_\even,\tilde T_\odd$ satisfy the same type of Lax equation  
\be  
\label{TBT}
\dot{\tilde{T}}_i=[ \tilde{B}_i, \tilde{T}_i], ~~~i=\even,\odd, \ee
with
\be \tilde{B}_\even={1\over 2}(\tilde{T}_\even^+-\tilde{T}_\even^-)+\dot\mu((\tilde{a}\tilde{a}^T)_\even^+-(\tilde{a} \tilde{a}^T)_\even^-), \quad \tilde B_\odd={1\over 2}(\tilde{T}_\odd^+-\tilde{T}_\odd^-).\ee
The relation between $T_i$ and $\tilde{T}_i$ in presence of $\mu\neq 0$ is the same as in the $\mu=0$ case considered earlier: 
they  are related  by an orthogonal transformation \eqref{Pevenodd}. To see that we note that since  $a$ is a normalized null vector of $T$,  we have $\tilde{a}=P_{\rm even}a$.

From \eqref{TBT} we can infer that the full Hamiltonian describes  two decoupled systems 
\be \mathcal H= \mathcal H_{\text{even}}+ \mathcal H_\text{odd}, \ee
where the odd subspace continues to evolve according to the usual Toda dynamics with the Hamiltonian
\be \mathcal H_\text{odd}=\mathcal I_2^\odd={1\over 2}\tr(T_\odd^2),\ee
and the even part is modified to
\be \mathcal H_{\text{even}}=\mathcal I_2^\even+\dot\mu\, \mathcal H',~~~\mathcal H'=\sum c_k\, \mathcal I^\even_k,  \label{todagen}\ee
where $c_k$ are some $\beta$-independent coefficients. This form is dictated by the fact that the set of all $\mathcal I^\even_k$ generate all possible isospectral deformations of $T_{\rm even}$ and hence describe \eqref{TBT} for some appropriate $c_k$. 
We find $c_k$ in terms of ${\mathcal I}^{\text{even}}_k$ by first bringing $T$ to the tri-diagonal form $\tilde{T}$ such that corresponding dynamical variables $\tilde{p}_n:= \dot{\tilde q}_n$ and $\tilde{q}_n$ have canonical Poisson brackets, and then comparing the resulting equations of motion $\dot{\tilde{T}}_{\rm even}=[ \tilde{B}_{\rm even}, \tilde{T}_{\rm even}]$ with $\dot{\xi}=\{H_{\text{even}},\xi\}$, for $\xi=\tilde{p}_n,\tilde{q}_n$ after imposing the constraint ${\rm det}(\tilde{T}_{\rm even})=0$. As a result we find 
\be  
\label{ck}
c_k =  \frac{2}{k!}\left. \frac{ \frac{\partial^k }{\partial \lambda^k } \det(\lambda\, \mathbb I - T_{\even} )}{ \frac{\partial }{\partial \lambda }   \det(\lambda\, \mathbb I - T_{\even} ) }\right|_{\lambda=0}.
\ee
Here the determinant of $T_{\even}$ is understood to be  a polynomial   of ${\mathcal I}^{\text{even}}_k$ for $1\leq k\leq (N+1)/2$, 
\be {\mathcal C}({\mathcal I}^{\text{even}}_k)=\det(T_{\even}). \ee 
This function  is   zero  ``on-shell'' because $T_{\even}$ has a zero mode.  
Then, as we show in the Appendix \ref{appB}, we can write 
\be 
\mathcal H' = 2 \left({\partial \ln \mathcal C\over \partial\, \mathcal I^{\text{even}}_1}\right)^{-1}.     \label{Hprime} \ee

\subsubsection{Example III}

Consider again the spin system introduced in section \ref{subsectionexample}.
\be H=-J(\sigma_x^1\sigma_y^2+\sigma_y^1\sigma_x^2)+g(\sigma_y^2+\sigma_x^1).\ee
Starting from the initial operator, which has non-zero diagonal elements in the energy eigenbasis,
\be A_0=\sigma_z^1\sigma_z^2,\ee
and performing Lanczos algorithm with $\beta_0=0$ we find
\begin{eqnarray} 
A_1&=&4ig(\sigma_z^1\sigma_x^2-\sigma_y^1\sigma_z^2),\\
A_2&=&8gJ(\sigma_x^1+\sigma_y^2)+16g^2\sigma_y^1\sigma_x^2,
\end{eqnarray}
along with the Lanczos coefficients
\begin{eqnarray}
    &b_0(0)=2\sqrt{2}|g|,\quad b_1(0)=2\sqrt{2g^2+J^2}.\label{bs2}
\end{eqnarray}
We also evaluate
\be h_i(0)=-{1\over 2}\tr(A_i^\dagger \{H,A_i\})/\tr(A_i^\dagger A_i),\ee
and
\be e^{2\mu(\beta,0)}\equiv\langle d,d \rangle_{\beta},\ee
resulting in
\be
     h_0(0)=\frac{2 g^2 J}{4 g^2+J^2},\quad h_1(0)=J,\quad
     h_2(0)=\frac{2 g^2 J+J^3}{4 g^2+J^2},\label{hs2}
\ee
\be e^{2\mu(\beta,0)}=\frac{2 J^2 e^{\beta  J} \cosh (\beta  \Delta )+\Delta ^2
   \left(e^{-2 \beta  J}+1\right)}{2 \left(\Delta ^2+J^2\right)},\quad \Delta\equiv \sqrt{4g^2+J^2}.\ee

The Krylov space is 3-dimensional. Let us write 
\begin{eqnarray}  
&&T^\text{even}=\begin{pmatrix}
	h_0 & t_{02}\\
	t_{02} & h_2 
\end{pmatrix},\qquad T^\text{odd}=\begin{pmatrix}
	h_1
\end{pmatrix},\\
&&t_{02}={b_0\over b_1}(h_1-h_0) = {b_0\over b_1}(h_2), \end{eqnarray}
where we used (\ref{Trec}) to determine the off-diagonal entry of matrix $T$ in terms of $b_i,h_i$.
The normalized null-vector of $T$ and $M$ is
\be a=\left({-b_1\over \sqrt{b_0^2+b_1^2}},0,{b_0\over \sqrt{b_0^2+b_1^2}}\right),\ee
leading to 
\be B_\text{even}=\left({1\over 2}t_{02}-\dot\mu{b_0b_1\over b_0^2+b_1^2}\right)\begin{pmatrix}
    0 & 1\\
    -1 & 0
\end{pmatrix} .\ee
The Lax equations $\dot T=[B,T]$, $\dot M=[B,M]$ give the following system of differential equations
\be \dot h_0=-\dot h_2={b_0^2\over b_1^2}(h_0-h_1)^2+\dot\mu {2b_0^2\over b_0^2+b_1^2}(h_0-h_1),~~\dot h_1=0,\ee
\be \dot b_0={1\over 2}b_0(h_1-h_0)-\dot\mu{b_0 b_1^2\over b_0^2+b_1^2},~\dot b_1={1\over 2}b_1(h_2-h_1)+\dot\mu{b_0^2 b_1\over b_0^2+b_1^2}.\ee
Taking advantage of the fact that the quantities $\tr(M^2)$ and $\tr((T^\text{even})^k)=\tr((T^\text{odd})^k)$ for $k=1,2$ are conserved, we find the general solution
parametrized by the constants $c,C,\lambda$
\begin{eqnarray}
h_0&=&c{ Ce^{c\beta-2\mu(\beta,0)}\over 1+Ce^{c\beta-2\mu(\beta,0)}},\\
h_1&=&c,\\
h_2&=&c{1\over 1+Ce^{c\beta-2\mu(\beta,0)}},\\
b_0^2&=&{\lambda^2C\over C+e^{2\mu(\beta,0)-c\beta}},\\
b_1^2&=&{\lambda^2\over 1+Ce^{c\beta-2\mu(\beta,0)}}.
\end{eqnarray}
Note that ${\rm det}\, T=0$ during the entire flow, as follows from $\det(T) = h_2 (h_0 - (b_0^2/b_1^2) h_2)$ and $\frac{h_0}{h_2} = \frac{b^2_0}{b^2_1}$.

Given the initial values $b_i(0),h_j(0)$ (\ref{bs2},\ref{hs2}), the integration constants can be evaluated to be
\be c=J,~\lambda=2\sqrt{4g^2+J^2},~C={2g^2\over J^2+2g^2} .\ee
This completes the derivation of $b_n(\beta)$.
To find the modified Hamiltonian governing the even part, we first calculate
\be \mathcal C=(\mathcal I^\even_1)^2/2-\mathcal I_2^\even.\ee
Using (\ref{ck}) we find that $c_1=2$ and $c_2=-2/J$, therefore the modified Hamiltonian governing the dynamics of the ``even'' part can be written as
\be \hat{\mathcal H}_{\even}=\mathcal I^\even_2+2\dot\mu (\mathcal I_1^\text{even}-\mathcal I_2^\text{even}/J).\label{Hp1} \ee
Using (\ref{Hprime}) we find
\be {\mathcal H}'={ ( \mathcal I^\even_1)^2-2 \mathcal I^\even_2\over \mathcal I^\even_1|_{os}}={1\over J}((\mathcal I^\even_1)^2-2\mathcal I^\even_2),\ee
and the equivalent Hamiltonian (generating the same equations of motion as \eqref{Hp1} provided the initial conditions are the same), is  
\be \mathcal H_{\even}=\mathcal I^\even_2+{\dot\mu\over J}((\mathcal I^\even_1)^2-2\mathcal I^\even_2).\ee

\subsection{Systems with gap degeneracies and generalizations}
\label{degeneracysection}
Now  we consider arbitrary systems with possible energy gap degeneracies. This consideration applies to all Hermitian initial operators and general inner product (\ref{definip}) with a real parameter $\beta$ where $\rho_1(\beta),\rho_2(\beta)$ are Hermitian, positive semi-definite operators that commute with the Hamiltonian. Let $J$ be the self-adjoint super-operator such that $e^{J\beta}A=\rho_1(\beta)A\rho_2(\beta)$ for any operator $A$. For the Wightman-ordered inner product we have $J=-{1\over 2}\{H,\cdot\}$.

We start by extending the Krylov space by additional orthonormal operators such that the extended space contains $e^{J\beta}A$ for any $\beta$.   Let $N$ denote the size of the original Krylov space and $N'\geq N$ the size of the extended Krylov space. There are different  ways to define a basis $\{O_n\}$ in the extended space. The resulting normalized operators $O_n$ with $n<N$ are fixed, but there is freedom in the choice of $O_n$ for $i\geq N$, stemming from the freedom to rotate $O_n$ for $n\geq N$ among themselves. One way to fix this ambiguity is  to lift all degeneracies by introducing a small parameter $\epsilon$, perform the Lanczos algorithm, normalize the operators and send $\epsilon\to 0$ in the end. Another way to fix the basis is to perform Lanczos algorithm with the operator $[H,\cdot]+\epsilon J$, and send $\epsilon\to 0$.

We define extended matrices $T',M'$ similar to (\ref{Jrepgen}), but with $i,j$ running through the entire extended Krylov space
\be T'_{ij}=\langle O_i,J O_j\rangle_{\beta}, ~~~0\leq i,j\leq N'-1.\label{defT}\ee
\be M'_{ij}=\langle O_i,[H, O_j]\rangle_{\beta}, ~~~0\leq i,j\leq N'-1.\label{defMM}\ee
The superoperator $J$ is mapping extended Krylov space onto itself, and therefore using the same steps as in section \ref{subsection1} we obtain the Lax equations
\be \dot T'=[B',T'],~~\dot M'=[B',M'],\quad B'={1\over 2}((T')^+-(T')^-).\label{laxTprime}\ee

Let us denote by $T_K,M_K$ the submatrices of $T'_{ij},M'_{ij}$ with $0\leq i,j\leq N-1$ and then write them in the block form
\be T'=\begin{pmatrix}T_{\text{{\tiny K}}} & t\\ t^T &  T_{\text{{\tiny E}}} \end{pmatrix}\qquad M'=\begin{pmatrix}
    M_\subK & 0\\
    0 & M_\text{\tiny E}
\end{pmatrix}. \label{submatrix} \ee
The matrix $M'$ is block diagonal, due to the termination condition $b_{N-2}=0$ of the Lanczos algorithm.
From (\ref{laxTprime}) we obtain the following equations for $T_{\text{{\tiny K}}}$ and $M_\subK$
\be 
\dot T_{\subK}=[B_{\subK},T_{\subK}]+tt^T,~~B={1\over 2}(T_{\text{{\tiny K}}}^{+}- T_{\subK}^{-}), \label{Tdegen} \ee
\be \dot M_\subK=[B_\subK,M_\subK]. \label{Mdegen} \ee
We can also write the equation for $T_\subK$ as follows
\be \label{generalcase}
\dot T_{\subK}=[B_{\subK},T_{\subK}]-T_{\subK}^2+Y,
\ee
where
\be Y_{ij}=\langle JO_i,JO_j\rangle.\ee
The equation \eqref{generalcase} governs $\beta$-evolution of $b_n$ and is defined solely in terms of the original (not extended) Krylov space. We illustrate \eqref{generalcase} using the simplest example of a  quantum harmonic oscillator in Appendix \ref{appOsc}.

\section{Extensions and applications}
\label{sec:app}
\subsection{Krylov bootstrap}
The differential equations governing the $\beta$-dependence of $b_n$, combined with the known properties of physical systems under consideration, can be leveraged to provide non-trivial consistency conditions. We call this idea ``Krylov bootstrap" and  illustrate its use in an example of a symmetrically ordered thermal 2pt function of primaries in a general 2d conformal field theory. For primaries with conformal dimension $\Delta$ and a CFT on a line the 2pt function is given by  \be
\label{2ptcft}
C_\beta(t) \propto \cosh(\pi t / \beta)^{-2\Delta}.
\ee 
The corresponding Lanczos coefficient are \cite{KrylovinCFT},
\be b^2_{n} = (n+1)(n+2 \Delta) \frac{\pi^2}{\beta^2},  \label{cftanswer}. \ee 
For convenience we can write this as $b^2_{n}=f^2(n) \ell^2(\beta)$ for some functions $f$ and $\ell$.

Since the one-point function on the line vanishes, we can set $\mu=0$ and the equations governing the $\beta$-dependence are \eqref{TandMdot}. Consistency with  $\dot{M} = \comm{B}{M}$ would require 
\be h_n = n \frac{2 \partial \log \ell(\beta) }{\partial \beta}  + h_0(\beta). \ee
The factorization of $b_n(\beta)$  imposes consistency conditions on $f(n)$. When $f(n)$ and $\ell(\beta)$ are as in $\eqref{cftanswer}$ the recursion expressions for $\eqref{Trec}$ can be solved to write for even $k>0$,  
\be 
T_{n, n+k} = 
-\frac{2 (-1)^{\frac{k}{2}+1} (\Delta  (k+1)+n)
}{\beta  \left(k^2-1\right)}  \sqrt{\frac{\Gamma (k+n+1) \Gamma (n+2 \Delta )}{\Gamma
   (n+1) \Gamma (k+n+2 \Delta )}}.
\ee 
Evaluating the right hand side of \eqref{lax2}, one finds it not to match the left hand side, even if freedom of choosing $h_0(\beta)$ is taken into account. 
Introducing non-zero $\dot{\mu}$ can not resolve this issue because there is no dependence on $\dot{\mu}$ in the odd subspace. This incompatibility calls for a non-zero matrix $t$ in \eqref{Tdegen}, which is necessarily implying the spectrum is degenerate. In other words, the explicit form of the Lanczos coefficients in 2d CFT \eqref{cftanswer} ``knows'' that the CFT Hamiltonian is degenerate, a conclusion which is not manifest at the level of 2pt function \eqref{2ptcft}. 

Our argument above relied on the explicit form of $f(n)$ and our ability to solve the recursive relation for $T_{n,n+k}$.
It would be interesting to extend it for any scale-invariant system, when $b_n(\beta)$ factorizes into $f(n)\ell(\beta)$ with $\ell(\beta)=\pi/\beta$ and arbitrary $f(n)$.

\subsection{Numerical evaluation of $C_\beta(t)$}

The system of nonlinear equations \eqref{TandMdot} cannot integrated analytically,  with an exception of very special or very small systems. An important application of our results would be to solve these equations and their generalization \eqref{lax2} numerically,  integrating starting from a particular value $\beta=\beta_0$.  
The initial data would include values $\{ q_{n}(\beta_0) \}$ and $\{ \dot{q}_{n}(\beta_0) \}$. This is equivalent to knowing  $b_n(\beta_0)$ and matrix elements $-\frac{1}{2} \langle O_n, \{H, O_n \} \rangle_{\beta_0}$. For spin chains and other physical systems with a finite local Hilbert space these values can be conveniently calculated numerically for $\beta_0=0$.  
We note, that  to integrate \eqref{lax2}   one also needs to know $\dot{\mu}(\beta)$ for all $\beta$. 

Once Lanczos coefficients are evaluated for a given $\beta$, the moments $\mu_{2n}$ can be computed by a variety of methods  to eventually evaluate matrix elements $(M(\beta)^n )_{00}$ up to  sufficiently large $n$. For finite moderately large  $N$, direct  exponentiation of $( e^{i M(\beta) t} )_{00}$ is feasible, resulting in $C_{\beta}(t)$ for the given $\beta$ and arbitrary $t$. 

We demonstrate the feasibility of this approach in case  a small-size chaotic transverse field Ising model
\bea
\label{IsingH}
H=-4 \sum_{i=1}^{L-1} S^x_i S^x_{i+1}+2 \sum_{i=1}^L (g\, S^x_i+h\, S^z_i),\qquad S^a_i={1\over 2}\sigma_a^i,
\eea
with $L=5$ spins and $h=1.05$, $g=0.8$. Our starting point was $\beta_0=0$, in which case evaluation of $q_n(0)$, $\dot{q}_n(0)$ can be done by first evaluating $A_n,b_n$ for $\beta=0$ using the conventional Lanczos algorithm \eqref{lanczos} and then evaluating ${\rm Tr}(A^{\dagger}_n\{H,A_n\})$. 
As the initial operator we take $A=2S^x_2$. To evaluate $\mu(\beta)$ using its definition \eqref{ddef} we rely on the exact diagonalization of $A$, but note that no matrix exponentiation is necessary. 

\begin{figure}
    \centering
    \includegraphics[width=0.55\linewidth]{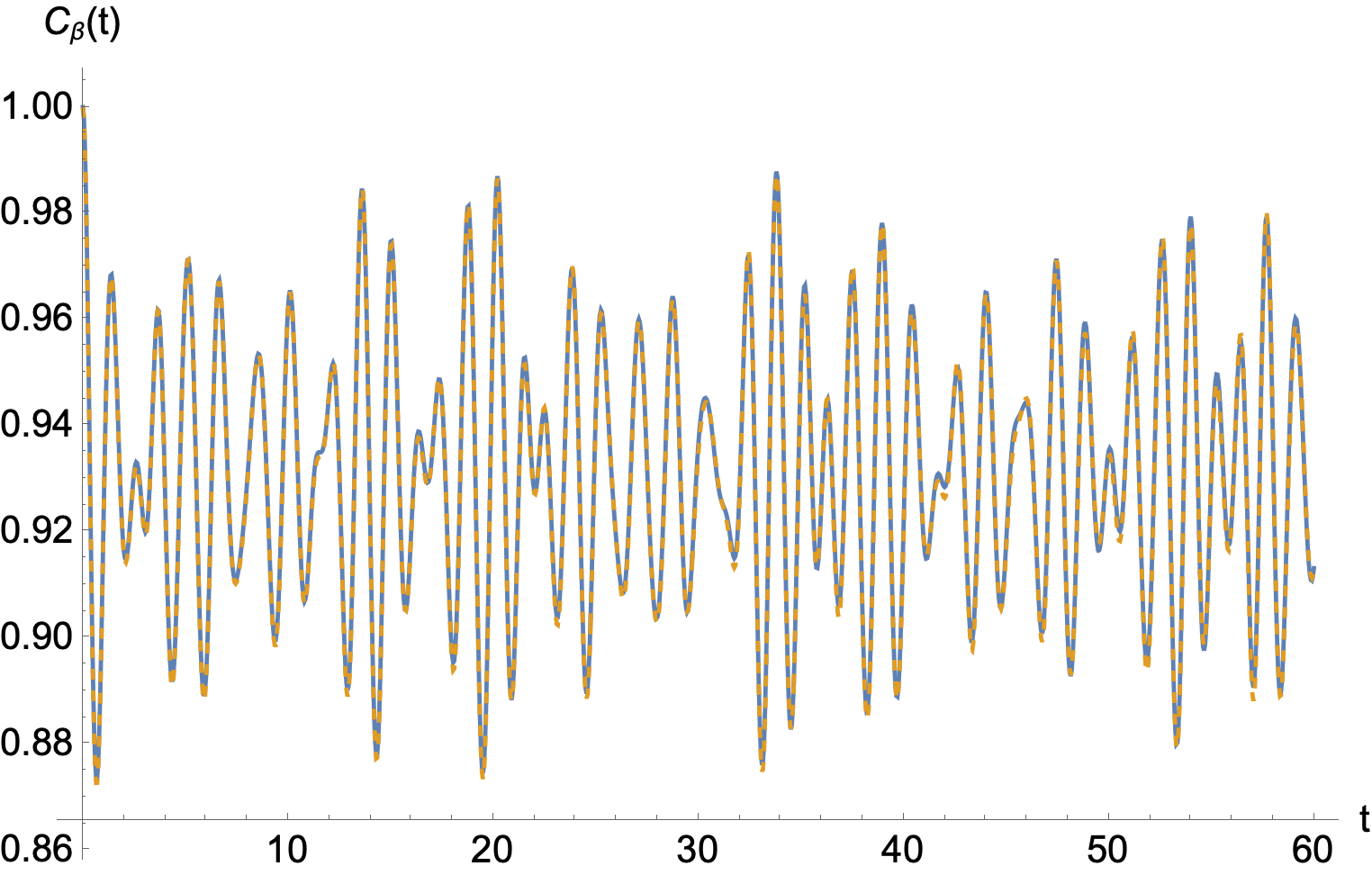}
    \caption{The autocorrelation function of $A=2S^2_x$ for the nonintegrable Ising model  \eqref{IsingH} with $h=1.05, g=0.8$ and $\beta=2$. Blue solid line: $C_\beta(t)$ obtained via  direct diagonalization of $H$ to evaluate $e^{-\beta H}$ and $e^{-i H t}$. Dashed orange line: $C_\beta(t)$ obtained via numerical integration of $b_n(\beta)$ using Euler algorithm and the subsequent evaluation of $(e^{i M(\beta) t} )_{00}$. In the latter case, the initial conditions were specified at $\beta_0=0$  and the size of the Euler step was $\Delta\beta=3.3\times 10^{-4}$.}
                                                \label{fig:numericalCtplot}
\end{figure}

To integrate \eqref{MTdynamics}  numerically to $\beta=2$  we used the  Euler algorithm with the step $\Delta\beta=3.3\times 10^{-4}$. 
We  observed empirically that the size of the discretization step  $\Delta\beta$ is not the primary factor affecting the final precision. Rather, values of $b_n(\beta)$ are very sensitive to the precision of the initial data $b_n(\beta_0)$ and $\dot{q}_{n}(\beta_0)$. 
We illustrate the result in Fig.~\ref{fig:numericalCtplot} where we plot $C_\beta(t)$ for $\beta=2$ and $t$ in the range $0\leq t\leq 60$ obtained in two ways, first via numerical integration of $b_n(\beta)$ and the subsequent exponentiation of $e^{i M(\beta) t}$ vs a straightforward approach of direct diagonalization.

\subsection{Large $\beta$ asymptotics of the Lanczos coefficients}
\label{asymptoticsec}
A well-known property of the Toda flow and its generalizations is that the corresponding matrices approach the diagonal form at late times \cite{Moser,TodaFlows}. A similar result holds for the $\beta$-evolution of $T$ and $M$ in the limit of $\beta m \gg 1$, where $m$ is the spectral gap defined as the size of the region where the spectral function $\Phi(\omega)$ (\ref{ipandauto}) vanishes,
\bea
\Phi(\omega)=0,\qquad 0<|\omega|\leq m.
\eea
The spectral gap $E_1-E_0$ is always equal to or smaller than $m$.

We  first consider the case when $\mu = 0$, i.e.~when the initial operator has no diagonal matrix elements in the energy eigenbasis. Equivalently, this is the case of an operator with vanishing thermal expectation value at any temperature. In this case, the eigenvalues of $T$ are the energy sums  $\lambda_{k} = -\frac{1}{2}( E_i + E_j)$ with each eigenvalue doubly degenerate. 

In this limit $T$ converges to a diagonal matrix $T_{jk} = \lambda_{k} \delta_{jk}$ with the eigenvalues $\lambda_k$ appearing in the descending order: $\lambda_0 = \lambda_1 > \lambda_2 = \lambda_3  \ldots$, where $\lambda_0 = -\frac{1}{2}(E_{0} + E_{1} )$. This is the only {\it stable} stationary fixed point of the flow. To see this, we first write the partial sum of $\dot{T}_{ii}$
\be 
\sum^{l}_{i=0} \dot{T}_{ii}   = \sum_{j\leq l, k > l } T^2_{jk} \label{partial_sum},
\ee 
that follows from  the Lax equation \eqref{TandMdot}. Assuming that the maximum eigenvalue of $T$ is well-defined (possibly using regularization) each $T_{ii} \leq \text{max}~ \lambda_{k}$ and thus the LHS of \eqref{partial_sum} is bounded from above. Since the RHS of \eqref{partial_sum} is positive semidefinite the only way for the LHS to be bounded at late times is if the RHS vanishes asymptotically. This must hold for all possible choices of $l$, from which we conclude that $T$ converges to a diagonal matrix. The ordering of the eigenvalues follows from the equation for $\dot{T}_{j,j+2}$. Shifting $T_{j,j+2} \rightarrow T_{j,j+2} + \epsilon(\beta)$ and treating $\epsilon$ as a perturbation gives
\be \dot{\epsilon} = \epsilon(\beta) (T_{i+2, i+2} - T_{i,i} ), \label{ep}\ee 
suggesting $\epsilon$ would grow unless $T_{i+2, i+2} \leq T_{ii}$.  

In the limit of large $\beta$, the  Lanczos coefficients are fixed by the fact that $T$ and $M$ commute, 
\begin{equation}
    \comm{T}{M}_{k,k+1} = b_k ( \lambda_{k} -\lambda_{k+1} )=0. 
\end{equation}
From here we conclude that Lanczos coefficients with odd index vanish $b_{2k+1}\rightarrow 0$. Thus matrix $M$ is block-diagonal, consisting of  $2\times 2$ blocks. Since $T$ and $M$ commute, matrices $T$ and $M^2$ share the eigenvectors, which implies that the Lanczos coefficient $b_{2k}$ with the same index as  $\lambda_{2k} = -\frac{1}{2}( E_i + E_j)$ is $b_{2k} = |E_{i} - E_{j}|$. 

When the initial operator has a non-zero thermal 1pt function, the spectrum of $T$ includes a zero eigenvalue, with all other eigenvalues being sums of the form $-{1\over 2}(E_i+E_j)$. Once again, $T$ asymptotes to the diagonal matrix of its eigenvalues. It is more convenient to consider the matrix $ \mathcal{T} = T +2 \dot{\mu} a a^{T}$, as in \eqref{relationbetweenTs}, that satisfies 
\be \dot{ \mathcal{T}  } =  [B,  \mathcal{T}] + 2 \ddot{\mu} a a^{T},\qquad  B = \frac{1}{2} (\mathcal{T}^{+} -  \mathcal{T}^{-} ). \label{shiftedTdot} \ee 
At large $\beta$ it will approach the diagonal form, with the same eigenvalues as $T$ except for  the zero eigenvalue that becomes 
\bea
\label{mulim}
\lim_{\beta \rightarrow \infty} 2\dot{\mu} = - E_{i_{*}}.
\eea
Here $E_{i_{*}}$ is the smallest energy for which $A_{i_{*}i_{*}} \neq 0$. 

To see that $\mathcal{T}$ approaches a diagonal matrix, we notice that in the RHS of \eqref{shiftedTdot} the partial sum along the diagonal is positive semidefinite. This is because the explicit expression for $\ddot{\mu}$ shows that it can be interpreted as  energy variance, hence $\ddot{\mu} \geq 0$. It can  be also checked that $\lim_{\beta \rightarrow \infty} \ddot{\mu} \rightarrow 0$ and thus, similarly to the argument above, we conclude that the diagonal elements of $\cal T$ become constant in the $\beta\rightarrow \infty$ limit. This, in particular, means that the vector $a$, which is an eigenvector of $\cal T$, takes the form $a_n = \delta_{n k^*}$, where $k^*$ is the location of the eigenvalue $-  E_{i_{*}}$ along the diagonal.

To see that the eigenvalues of $\cal T$ will be arranged in descending order, it is enough to repeat the argument around \eqref{ep}.

As in the case without $\mu$, the asymptotic value  of Lanczos coefficients can be determined using that $M$ and $\cal T$ commute. In the large $\beta$ limit, in addition to $2 \times 2$ blocks, $M$ will have a $1 \times 1$ block with zero eigenvalue located at $k^*$-th place, with the corresponding eigenvector given by $a$. The two adjacent Lanczos coefficients will vanish as well, $b_{k^*-1}=b_{k^*}=0$. 
Assuming for simplicity that the operator $A$ has a non-zero vacuum expectation value $A_{00} \neq 0$, such that $E_{i_{*}}$ is the ground state energy, we readily conclude that the eigenvalue of $\cal T$ associated with $a$ will appear first, $k^*=0$. In this case 
all even Lanczos coefficient will vanish $b_{2k} \rightarrow0$, while 
the odd ones are given by $b_{2k+1} = \abs{E_i-E_j}$,  where $E_i,E_j$ are determined such that corresponding eigenvalue of $\cal T$ is $\lambda_{2k+1} = -\frac{1}{2}(E_i + E_j)$.

\subsection{K-complexity in the very large $\beta$ limit} 
\label{Kcomplexity}
Finally, we discuss the behavior of Krylov complexity when $\beta$ is large. Krylov complexity is defined as the 
mean position of the operator $O(t)$ spread along the ``Krylov chain,''
\be 
K(t) = \sum^{n=N-1}_{n=0} n |\langle O_{n}, O(t) \rangle_{\beta}|^2 . \label{defKt}
\ee
This can be written as $K(t) = \sum^{n=N-1}_{n=0} n \phi_n^2(t)$, 
where 
\be
O(t)=\sum_n i^n \phi_n(t) O_n,
\ee
and $\phi_n(t)$ satisfy 
\begin{eqnarray}
&&\dot{\phi}_n(t) = b_{n-1} \phi_{n-1}(t)+b_n \phi_{n+1}(t)  \label{discreteSE},\\
&&\phi_n(0)=\delta_{n,0}.
\end{eqnarray}

\begin{figure}    
\centering
         \includegraphics[width=0.5\textwidth]{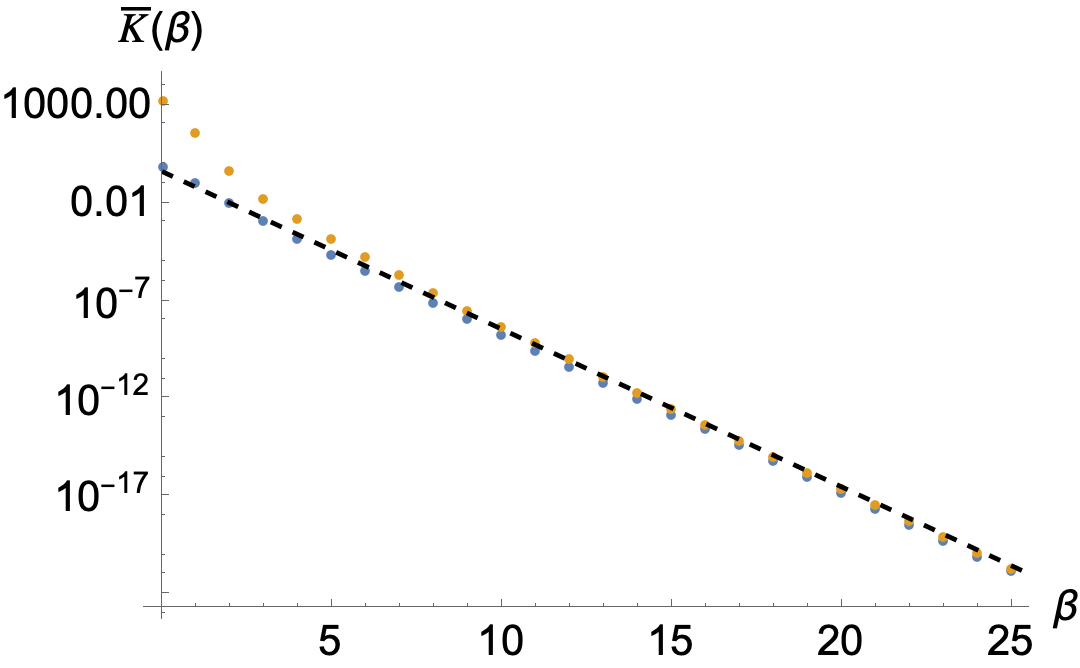}
\caption{ Time averaged Krylov complexity $\overline{K}(\beta)$ of $A=S^x_2$ in the chaotic Ising spin chain \eqref{IsingH} with $L=6,g=0.8,h=1.05$. We show  $\overline{K}(\beta)$ evaluated using direct diagonalization (blue dots) vs the approximation  in terms of the first few $b_n$ \eqref{kbar_approx_formula}. Dashed black line is $ \# e^{-\beta m/2}$ with $\#$ a fitting constant.}
   \label{Kbar}
\end{figure}

Normally $K(t)$ grows with time then oscillates around the values of order the length of the chain $e^S$. But when temperature is extremely small, half of the Lanczos coefficients effectively vanish, and the operator is getting confined at the first few sites. To model that behavior we consider the case of an operator with non-zero thermal expectation value, such that $b_{2n}\rightarrow 0$. After taking $b_2\rightarrow 0$ the dynamics can be solved explicitly 
\begin{eqnarray}
&&\phi_0(t) = \frac{b_0^2 \cos \left(\sqrt{b_0^2+b_1^2}
   t\right)+b_1^2}{b_0^2+b_1^2},\\
&&\phi_1(t) = \frac{b_0 \sin \left(\sqrt{b_0^2+b_1^2}
   t\right)}{\sqrt{b_0^2+b_1^2}},\\
&& \phi_2(t) = 
\frac{b_0 b_1 \left(\cos \left(\sqrt{b_0^2+b_1^2}
   t\right)-1\right)}{b_0^2+b_1^2}.
\end{eqnarray}
After evaluating the time average we find 
\be 
\overline{K}(\beta) \approx |\overline{\phi}_1|^2 +2 |\overline{\phi}_2|^2  \approx \frac{b_0^2 \left(b_0^2+7 b_1^2\right)}{2
   \left(b_0^2+b_1^2\right){}^2}. \label{kbar_approx_formula}
\ee 
Taking into account that $b^2_0 \approx e^{-\frac{\beta  m}{2}}$ is exponentially small when  $\beta m \gg 1$, we readily find that the averaged Krylov complexity decreases exponentially with $\beta$, 
\be
\ln\overline{K}(\beta) \approx -\beta m. 
\ee
This is the behavior observed numerically, as shown in Fig.~\ref{Kbar}.

\section{Moderate $\beta$ regime}\label{sec:moderate}
In the section above we discussed the regime of asymptotically large $\beta$, when $\beta m\gg 1$ and the asymptotic behavior can be understood analytically. In this section we discuss the regime of moderately large $\beta m \gtrsim  1$, and describe the typical behavior of $b_n$  qualitatively.

We start with the spectral representation of the Wightman function
\be  \label{Phi}
C_{\beta}(t) = \int ~ e^{i (E_{1} - E_{2} ) t}  e^{- \beta \frac{E_{1} +E_2}{2} }  \rho(E_1, E_2) |\bra{E_1} O \ket{E_2}|^2 dE_{1} dE_{2},  \ee
where $\rho(E_1, E_2)$ is the joint density of eigenvalues. It is useful to change variables to $E=\frac{1}{2} (E_1 + E_2)$ and $\omega = E_1 - E_2$ so that 
\be  C_{\beta}(t) = \int ( g(\omega, \beta)  + \kappa(\beta) \delta(\omega) )   e^{i \omega t}  d\omega =\int  ~  \Phi(\omega) e^{i \omega t} ~d\omega. \label{measuredef} \ee 
Here
\be g(\omega, \beta) = \int  e^{-\beta E} \rho\big(E+\frac{\omega}{2} \big)\rho\big(E-\frac{\omega}{2} \big) |O(E, \omega)|^2  dE ~~ , ~~ \kappa(\beta)= \int   e^{-\beta E} \rho(E) |O(E, 0)|^2 dE,  \label{kgdef} \ee
and we used $O(E,\omega)=\langle E_1|O|E_2\rangle$ without assuming that it is a smooth function of its arguments. 

Non-zero $\kappa$ indicates $O$ has non-zero thermal expectation value. When $\beta$ is large, provided the system exhibits a spectral gap $m$ or $O(E,\omega)$ vanishes for small non-zero $|\omega|<m$ for some $m$, one can approximate \eqref{measuredef} as follows
\be
\Phi(\omega) = \begin{cases}
			\frac{1}{\mathcal N} e^{- \beta \abs{\omega}/2} + \kappa(\beta)  \delta(\omega), &  \text{if} ~ m < \abs{ \omega } < \omega_{\max}, \\
            0, &   \text{otherwise}.  \label{toymeasure}
		 \end{cases}
\ee 
Here $\mathcal N$ is the normalization factor 
\be 
\mathcal N = \frac{4 
   \left(e^{ -  \frac{1}{2} \beta m  }-e^{- \frac{1}{2} \beta \omega_{\max} 
   }\right)}{\beta }.
\ee 
This is a simple model that captures the typical behavior of $b_n$  when $\beta m \gtrsim  1$. The cutoff $\omega_{\max}$ is  introduced to model the UV-effects when $b_n$ saturate to a constant. 

For simplicity we start with $\kappa=0$. A simple calculation for the moments $\mu_n$ yields (all odd moments vanish)
\be
\mu_{2n} = \frac{2}{\mathcal N } \Big( \frac{2}{\beta} \Big)^{2 n+1} 
   \left( \Gamma \left(2
   n+1,\frac{\beta  m
   }{2}\right) - \Gamma \left(2
   n +1,\frac{\beta \omega_{\max} 
   }{2}\right)  \right).
\ee 
First, we take $m=0$ and for $2n \lesssim \frac{1}{2} \beta \omega_{\max}$, the cutoff can be taken to infinity, 
\be
\mu_{2n} \approx \begin{cases}
			\frac{2}{\mathcal N} 2n! \Big( \frac{2}{\beta} \Big)^{2 n+1}  , &  \text{for} ~2n \lesssim \beta \omega_{\max}/2, \\
        \frac{2}{\mathcal N}  \frac{ \omega^{2n+1}_{\max}  }{ 2n} e^{ - \frac{1}{2} \beta \omega_{\max}},  &   \text{for} ~ 2n \gtrsim  \beta \omega_{\max}/2.
	 \end{cases}  \label{approxmoments}
\ee 
From here we find that Lanczos coefficients behave as follows
\begin{eqnarray}\nonumber
b_{n} \approx \frac{\pi}{\beta}n,\qquad  n\lesssim  n^{*},\\
b_{n} \approx \frac{\pi}{\beta} n^{*},\qquad n \gtrsim  n^{*}, \label{bbehavior}
\end{eqnarray}
where $n^{*} =  \frac{\beta\, \omega_{\max} }{2 \pi} $. This behavior is confirmed 
by the numerical examples discussed below. 


Next, we consider $m> 0$. Whenever $\Phi(\omega)$ exhibits a gap, a general mathematical result of \cite{Chihara} suggests that Lanczos coefficients split into even and odd branches, with the asymptotic behavior 
\be 
\label{asympt}
\abs{b_e - b_o} = m  ~~\text{and}~~ b_e + b_o =\omega_{\max}.
\ee 
Here $b_e=\lim_{n\rightarrow \infty} b_{2n}$ and $b_o=\lim_{n\rightarrow \infty} b_{2n+1}$ are assumed to be finite, which is always the case when the spectral support of $\Phi(\omega)$ is bounded from above. Second equation in \eqref{asympt} can be also derived with help of the Dyck path formalism, see Appendix \ref{Dyck}.

\begin{figure}  
\centering
    \begin{subfigure}[h]{0.46\textwidth}
    \centering
    \includegraphics[width=\linewidth]{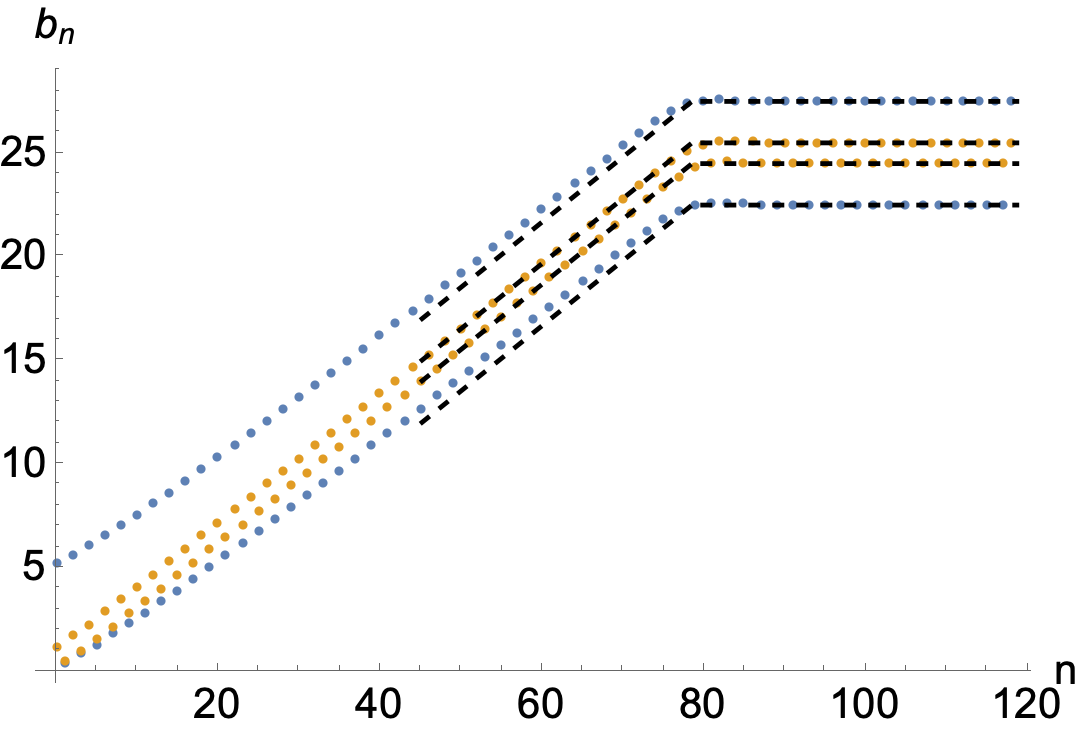}
        \caption{}
    \label{fig: numexample}
    \end{subfigure}
     \begin{subfigure}[h]{0.49\textwidth}
      \includegraphics[width=\linewidth]{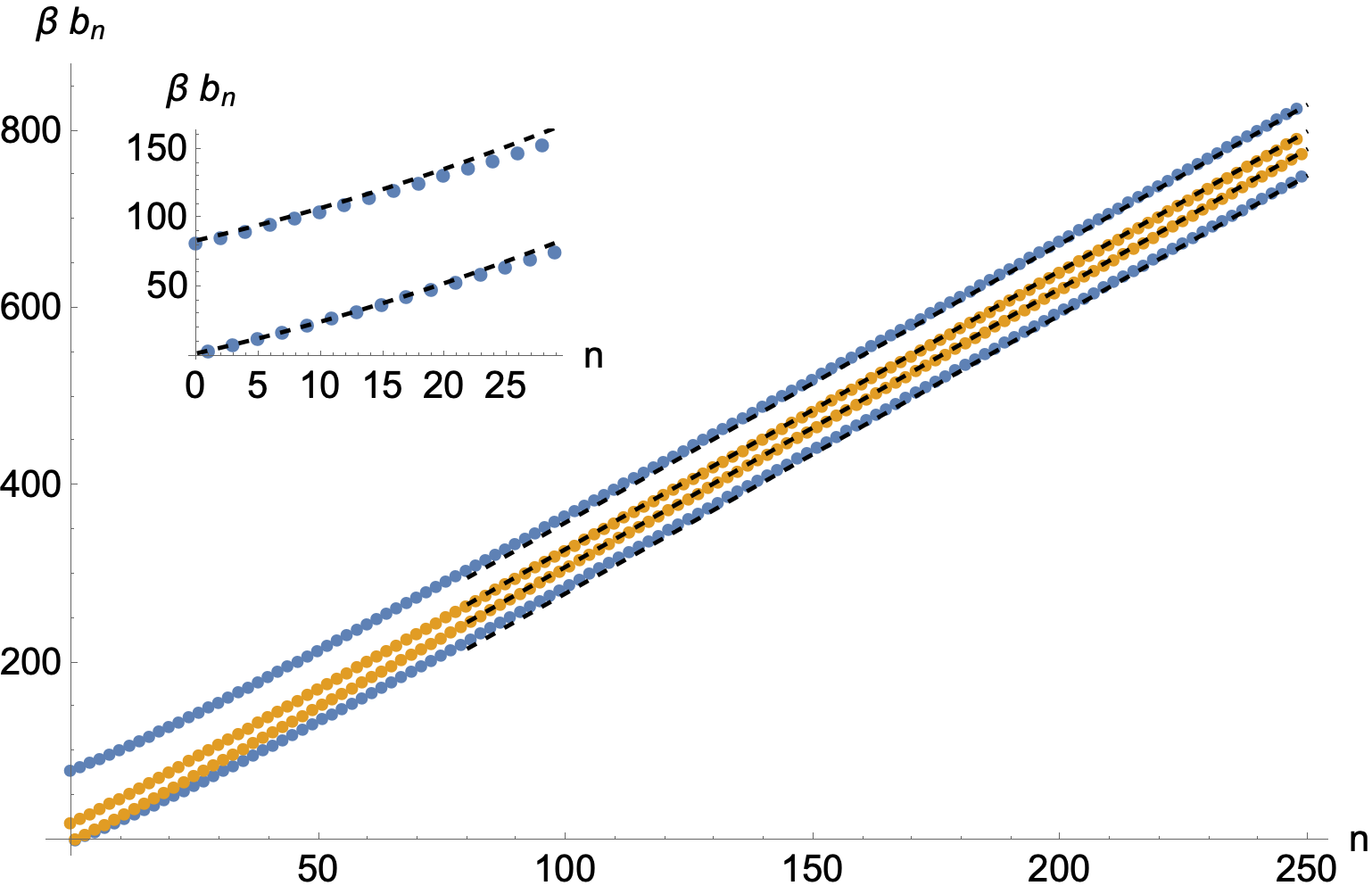}
      \caption{}
       \label{d6plot}
\end{subfigure}
\caption{
In \ref{fig: numexample}, we show Lanczos coefficients $b_n$ for the model \eqref{toymeasure} with $m=1$ (orange) and $m=5$ (blue) for $\beta=5$.
Black dotted line show the approximation \eqref{asympt} for $n\geq n^*$ and \eqref{bgrowth}, \eqref{asymptC} for $n<n^*$, where $n^*= \beta \omega_{\text{max}}/(2\pi)$. A good agreement with exact Lanczos coefficients confirms that $|b_{\even}(n)-b_{\odd}(n)|\approx m$ well before $b_n$ saturate to their UV values. 
In \ref{d6plot} we plot  Lanczos coefficients $b_n$ for free massive scalar field theory in 6d with $\beta m=20$ (orange) and $\beta m=80$ (blue). The black dashed lines are fits given by  \eqref{bgrowth} with the respective values of $\beta$, $m$ and $\Delta$.  The inset shows $b_n$ for $\beta m=80$ for small $n$ (blue) superimposed with the approximation \eqref{smallnfieldtheory} (dashed black line). 
 }
\end{figure}


We illustrate \eqref{asympt} with a simple model \eqref{toymeasure} with $m > 0$, which yields a generalization of (\ref{approxmoments}), 
\be
\mu_{2n} \approx \begin{cases}
			\frac{ m^{2j} e^{-\frac{\beta}{2} m}}{m}  \frac{2}{\mathcal N}, &  \text{for} ~ 2n \lesssim \beta m/2, \\
			\frac{2}{\mathcal N} 2n! \Big( \frac{2}{\beta} \Big)^{2 n+1},  &  \text{for} ~ \beta m/2\lesssim 2n \lesssim  \beta\omega_{\max}/2, \\
           \frac{ \omega^{2n+1}_{\max}  }{ 2n} e^{ - \frac{1}{2} \beta \omega_{\max}} \frac{2}{\mathcal N}, &   \text{for} ~ 2n \gtrsim \beta \omega_{\max}/2.
	 \end{cases}  \label{approxmoments2}
\ee 


We plot the corresponding $b_n$ in Fig.~\ref{fig: numexample}.
As can be seen there, the distance between even and odd branches remains approximately the same $|b_{\even}(n)-b_{\odd}(n)|\approx m$  for large $n$ well before linear growth saturates for $n\geq n^{*} =  \frac{\beta\, \omega_{\max} }{2 \pi} $. (Here $b_{\even}, b_{\odd}(n)$ represent smooth behavior of $b_n$.) 

In cases, when there is only one smooth branch of $b_n$, in the regime of asymptotic linear growth, i.e.~for $n\ll n^*$, the slope and the intercept are determined by the location and the order of the singularity of $C_\beta(t)$ along the imaginary axis \cite{KrylovinCFT}. When $b_n$ split into two branches due to $m>0$, in the simplest scenario the slope remains the same while each branch has its own intercept
\be
\label{bgrowth}
b_n\approx {\pi n\over \beta} +{c_o+c_e\over 2}+(-1)^n{c_e-c_o\over 2}+\dots 
\ee
Here we implicitly assume that $C_\beta(t)$ diverges at $t=\pm i\beta/2$, consistent with \eqref{toymeasure} in the $\omega_{\text{max}}\rightarrow \infty$ limit. 
The order of the singularity of $C_\beta(t)$ controls ${c_o+c_e}$ but not ${c_e-c_o}$  \cite{avdoshkin2022krylov},
\be
{c_e+c_o}= \frac{\pi}{\beta}(2\Delta-1),  \label{ceplusco}
\ee
where $\Phi(\omega)=\omega^{2\Delta-1}e^{-\beta \omega/2}$ for large $\omega$, or equivalently $C_\beta(t)\propto |t-i\beta/2|^{-2\Delta}$ when $t\rightarrow i\beta/2$. 
Together with \eqref{asympt} and the observation that the difference $|b_{\text{even}}(n)-b_{\text{odd}}(n)|\approx m$ remains approximately the same for $n\gg 1$ this gives
\be
\label{asymptC}
c_o={1\over 2} \big( {\pi \over \beta}(2 \Delta-1) + m \big)  ,\qquad 
c_e={1\over 2}  \big( {\pi \over \beta}(2 \Delta-1) - m\big).
\ee
This expression clearly shows that the intercepts know about both UV and IR behavior of $\Phi(\omega)$. 

For the moments \eqref{approxmoments} and small $n$ an explicit calculation yields 
\begin{eqnarray}\nonumber
b_{2n}  &=& m+{2n\over \beta}+{(2n)^2\over 2\beta^2 m}+O \Bigl(  \frac{1}{\beta^3 m^2 },  \Bigr)\\ 
   b_{2n+1} &=& {2n+1\over \beta}+{(2n+1)^2\over 2\beta^2 m}+O \Bigl( \frac{1}{\beta^3 m^2 } \Bigr).  \label{smallnfieldtheory}
\end{eqnarray}
These expressions can be trusted so far $n\ll \beta m$, as we illustrate using a numerical example below. 

Comparing \eqref{smallnfieldtheory} with \eqref{asymptC} we conclude that in the presence of $m>0$, Lanczos coefficients $b_n$ split into two smooth branches, both initially growing linearly and saturating around the cutoff $b_n \approx \omega_{\max}/2$. The intercepts of the two branches are initially $m$ and $0$ and then shift gradually to \eqref{asymptC} for $n\lesssim n^*$ and \eqref{asympt} for $n\gtrsim n^*$. 

We illustrate the behavior outlined above using free massive scalar theory in $d=6$, see \cite{KrylovinCFT,avdoshkin2022krylov}. We plot corresponding Lanczos coefficient in Fig.~\ref{d6plot} for different scalar mass $m$, which is also the spectral gap of $\Phi(\omega)$. Comparing with the asymptotic behavior of (\ref{bgrowth},\ref{asymptC}) we  see that $b_e-b_o$ quickly saturates to $m$ and remains approximately constant.  The inset in Fig.~\ref{d6plot} confirms \eqref{smallnfieldtheory} is a good approximation for large $\beta$ and small $n$.

\begin{figure}  
\centering
\includegraphics[width=0.5\linewidth]{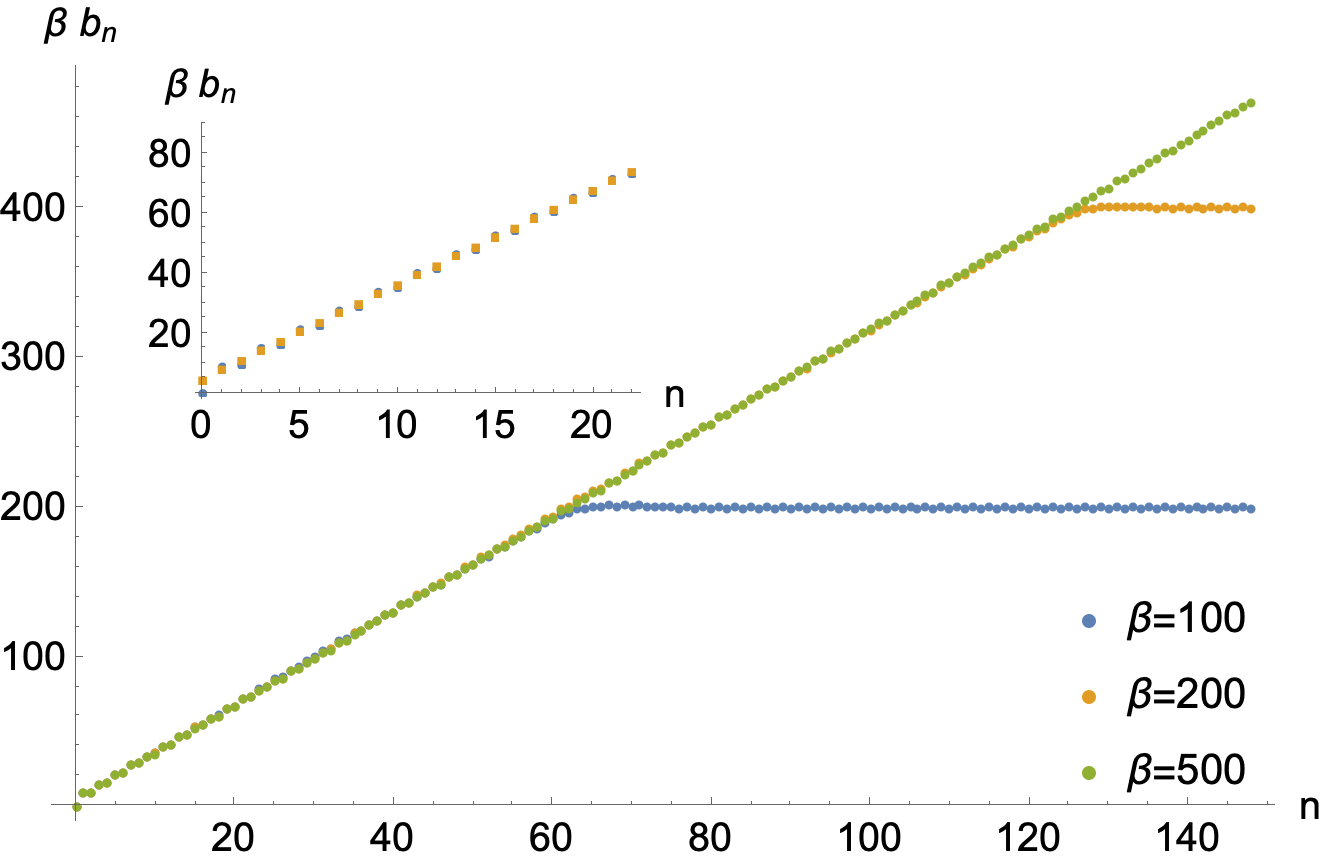}
\caption{Lanczos coefficient  $b_n$ for critical Ising model  \eqref{XYchain} with $h=1$, $\gamma=1$. The increasing range of linear growth with $\beta$ illustrates \eqref{bbehavior}. The inset compares the Lanczos coefficients (blue) with the approximation \eqref{tildeb}   for $\beta=500$ (orange).   }
\label{kappaIsing}
\end{figure}

Finally, we return to the model \eqref{toymeasure} and turn on positive $\kappa$, which means $C_\beta(t)$ does not vanish as $t\rightarrow \infty$. The new correlation function is
\be
\tilde{C}_\beta(t)=C_\beta(t)+\kappa\, C_\beta(0),
\ee
where we introduced a constant $C_\beta(0)$  for convenience. 
The relation between  Lanczos coefficients $\tilde{b}_n$ for $\tilde{C}_\beta$ and $b_n$ of $C_\beta(t)$ are given by a  known recursive relation \cite{UVAROV196925}\footnote{A related iterative relation between $\tilde{b}_n$ and $b_n$ was recently discussed in  \cite{Gamayun:2025hvu}.}
\begin{equation}
\label{delta_recursion1}
b^2_{2n+1} b^2_{2n+2} = \tilde{b}^2_{2n+1} \tilde{b}^2_{2n+2}, \quad \text{for} \quad n= 0,1,2\dots,  
\end{equation}
and 
\begin{equation}
\label{delta_recursion2}
b^2_{2n} + b^2_{2n+1} = \tilde{b}^2_{2n} + \tilde{b}^2_{2n+1}, \quad \text{for} \quad n= 0,1,2\dots 
\end{equation}
This relation is fully general whenever $C(t)$  is deformed by an additive constant. 
The sequences $\tilde{b}_{n}$ is fully determined by \eqref{delta_recursion1}  and \eqref{delta_recursion2} and the value of $\tilde{b}_0$. The latter is $\kappa$-dependent and is given by
\begin{equation}
    \tilde{b}_0^2 =\frac{b_0^2}{1+\kappa},\qquad {b}_0^2 =\frac{\int \omega^2 \Phi(\omega) d\omega }{\int \Phi(\omega) d\omega }.
\end{equation}
If the sequence $b_n$ has smooth dependence on $n$, i.e.~there is only one smooth branch $b_n^2=b^2(n)$, then the sequence $\tilde{b}_{n}$ will split into two smooth branches for even and odd $n$, exhibiting staggering. This follows from \eqref{delta_recursion1} and \eqref{delta_recursion2} as we now explain. To see that, we notice that introducing non-trivial $\kappa$ does not change the odd Krylov vectors $A_{2n+1}$, as follows from the recursion relation \eqref{lanczos}. The norm of $A_{2n}$ increases when $\kappa>0$ because of the  diagonal matrix elements of $A_{2n}$ in the energy eigenbasis. Given that $b_n^2$ is the ratio of norms of $A_{n+1}$ and $A_n$ \eqref{b2def}, we immediately conclude that 
 $b^2_{2n+1}$ will increase while $b^2_{2n}$ will decrease for $\kappa>0$. Because this deformation does not change the moments (except for $\mu_0$) the asymptotic behavior of $b_n$ will remain unchanged. 
 
 We illustrate the onset of staggering due to $\kappa$ with the help of a simple analytic example with all  $b_n = b$ being constant. This corresponds to the semicircle distribution $\Phi(\omega)=\frac{1}{2\pi b}\sqrt{4 b^2-\omega^2}$. Then  $\tilde{b}_{n}$ 
 associated with $\tilde{\Phi}(\omega)=\Phi(\omega)+\kappa \delta(\omega)$  
 can be found explicitly to be 
\be  
\tilde{b}^2_{n} = b^2  \frac{1+  \big( \lceil \frac{n+1}{2} \rceil - (-1)^n  \big) \kappa }{1+  \lceil \frac{n+1}{2} \rceil \kappa }.
\ee 
As expected,  for large $n$,  $\tilde{b}^2_{n}$ approaches $b^2_{n} = b^2$. 

The same  conclusion, that introduction of $\kappa>0$ in full generality leads to staggering, is supported by the linearized analysis. Starting from \eqref{delta_recursion1} and \eqref{delta_recursion2} and expanding at linear order in $\kappa$ yields 
\bea
\label{tildeb}
\tilde{b}^2_{2n} &=&b_{2n}^2 - \kappa \varepsilon_{2n} + O(\kappa^2) ,\\
\tilde{b}^2_{2n+1} &=&b_{2n+1}^2 + \kappa \varepsilon_{2n}+O(\kappa^2),\\
\varepsilon_{2n}&=&{\prod_{i=0}^n b_{2i}^2\over \prod_{i=0}^{n-1} b_{2i+1}^2}.
\eea
A similar quantity $\mathcal{A}_n ={\varepsilon_{2n}\over b_0 b_{2n}}$ was recently introduced in \cite{Bartsch_2024}, where it was shown that in the large $n$ limit it converges to the integral $\int_0^\infty dt\, C(t)$, provided it is finite. We thus immediately conclude that at least to leading order in $\kappa$, $\tilde{b}_{2n} \approx  b_{2n} - \frac{\kappa}{2} \frac{\varepsilon_{2n}}{b_{2n}}$ gets shifted only by a constant $\frac{\kappa}{2} \mathcal{A}_n b_0 $, with the same conclusion (and opposite sign) holding true for $b_{2n+1}$.

We illustrate the effect of $\kappa>0$ using our second example,  the XY model, a spin-chain with the Hamiltonian 
\be H=\sum_{j=1}^N[(1+\gamma)S_j^xS_{j+1}^x+(1-\gamma)S^y_j S^y_{j+1}]-h\sum_{j=1}^N S_j^z, \label{XYchain}\ee
considered in the thermodynamic limit $N\rightarrow \infty$. We will consider a one-site operator $A=S_z$, see Appendix \ref{XY} for details. 
Taking $h=\gamma=0$, one finds gapless behavior with $m=0$ and $C_\beta(t)\rightarrow 0$ for large $t$. In this case coefficients $b_n$ are well-modeled by \eqref{approxmoments} and \eqref{bbehavior} as was confirmed numerically in \cite{avdoshkin2022krylov}. Taking $h=\gamma=1$ will yield critical Ising model without mass gap but a non-zero asymptotic value of  $C_\beta(t)$ when $t\rightarrow \infty$, indicating $\kappa>0$. At long distances, this system is described by the Ising CFT. If the asymptotic value of $C_\beta(t)$ is subtracted from $C_\beta(t)$ then Lanczos coefficients would be given by $b_n \approx \frac{\pi}{\beta}( n + \Delta+ \frac{1}{2} )$. The actual value of $\kappa$ in this case is given by  $\kappa = m^2_z$, where $m_z(\beta)$ is the magnetization defined in \eqref{magdef}. This allows us to compute \eqref{tildeb} explicitly and find that the linearized answer well approximates the behavior of $b_n$ as shown in Fig.~\ref{kappaIsing}.


To summarize, we have identified two factors that  contribute to $b_n$ splitting into two smooth branches, that grow ``parallel'' to each other -- the behavior dubbed ``staggering'' in \cite{Mitra}. First is the spectral gap $m$, i.e.~vanishing of the spectral density $\Phi(\omega)$ for small frequencies $0<|\omega|\leq m$. Second is the delta-function at the origin $\omega=0$ of $\Phi(\omega)$, which is the same as the shift of $C(t)$ by a constant. As $\beta$ increases both factors become more pronounced. This is because large $\beta$ suppresses the large $\omega$ tail of $\Phi(\omega)$ by the factor $e^{-\beta \omega/2}$ as can be seen from  \eqref{Phi}, effectively zooming on the behavior near the origin. This explains why in a typical physical system, e.g.~those considered in \cite{Mitra,Mitra2,Mitra3,Yates:2021lrt,Yeh:2023fek,avdoshkin2022krylov,Camargo_2023,Tang:2023ocr,  tan2024scalingrelationsspectrumform}, Lanczos coefficients  exhibit linear growth and may split into two branches in the low energy regime.  

\begin{figure}
    \centering
    \includegraphics[width=0.5\linewidth]{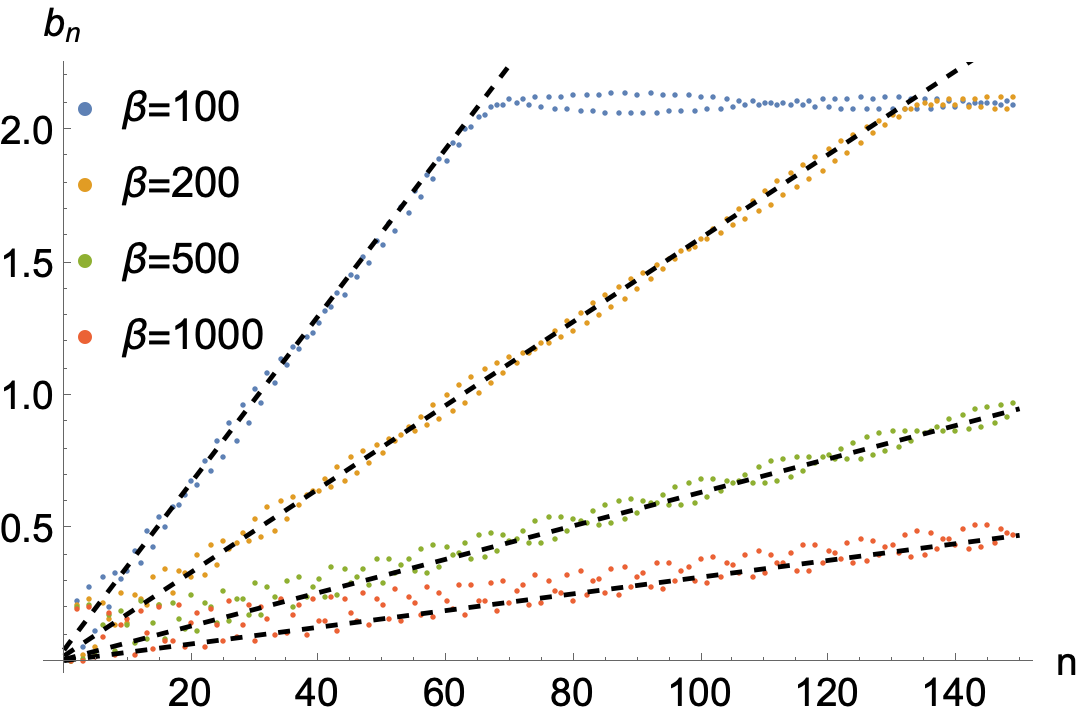}
    \caption{Lanczos coefficients for $XY$ model \eqref{XYchain} with $h=1.1, \gamma=1$ with mass $m=0.1$ for various $\beta$. Black dashed lines: approximate mean behavior as given by \eqref{bbehavior}.}
    \label{massiveXY}
\end{figure}

While the simple model \eqref{toymeasure} captures many qualitative features of $b_n$ for a typical physical correlator  at small temperatures, the behavior of $b_n$ discussed above, in particular staggering, is not universal. Additional features are present in physically relevant cases. For example, free massless field theories on spheres considered in \cite{avdoshkin2022krylov} give rise to $b_n$ splitting into two branches that grow with different {\it slopes}. Another interesting example is given by the XY model with general values of $h$ and $\gamma$. Consider e.g.~the model with $h=1.1,\gamma=1$ in the large-$\beta$ regime. In this case the spectral density is approximately given by \eqref{toymeasure} with both $m,\kappa>0$. And while {\it mean} growth of $b_n$ is still accurately captured by \eqref{bbehavior}, and $b_n$ split into even and odd branches that behave smoothly, the rest of the analysis from above does not apply. Instead of staggering, i.e.~growth parallel to each other, these branches exhibit ``oscillatory'' behavior interchanging places. This is shown in Fig.~\ref{massiveXY}. It would be interesting to develop a better analytic understanding of what causes this behavior and how to qualitatively describe it.          


\section{Discussion}
\label{sec:discussion}

In the paper we studied temperature dependence of Lanczos coefficients associated with the thermal 2pt function of a Hermitian  operator. First, we have shown that the temperature dependence is governed by a completely integrable system of equations related to the Toda hierarchy. For a chaotic system without energy gap degeneracies and for an  operator with vanishing thermal expectation value, this dynamics is described by two  independent Toda chains, related at the level of the initial conditions. For a more general systems and/or operators, one of the Toda chains is getting modified. 

Using known analytic behavior of the Toda systems at late times, we deduced a universal behavior of the Lanczos coefficients at very low temperatures $\beta \to \infty$, valid when $\beta$ is much larger than the inverse spectral gap of the system $m^{-1}$. In this limit half of the Lanczos coefficients vanish, while the other half converge to the values of the energy gaps.
The convergence is faster for $b_n$ with smaller value of the index. In the same limit time-averaged Krylov complexity becomes exponentially small, of order of $\bar{K}(\beta) \sim e^{-{\beta  m/ 2}}$. This value seems to be independent of the total system size.  

We have also investigated the behavior of Lanczos coefficients $b_n$ in the regime of moderately large $\beta \sim m^{-1}$. In this regime the behavior is not universal, but for many physical systems $b_n$ split into two smooth branches, for even and odd $n$. This behavior, when the two branches ascend with $n$ essentially parallel to each other, is called staggering in the literature \cite{Mitra}. We have identified two distinct mechanisms that cause staggering: spectral gap of the spectral measure $\Phi(\omega)$ (\ref{measuredef}) and the constant part of the 2pt function $C_\beta(t)$. As the temperature decreases, the effects of each of them becomes more pronounced. Using a simple model for $\Phi(\omega)$ we have analytically deduced typical behavior of $b_n$ exhibited by various systems, including spin chains or field theories, at low temperatures. Yet, there are also physical systems that demonstrate a more nuanced behavior of $b_n$ that goes beyond the aforementioned typicality. 

Our work opens several research directions, that we have preliminarily investigated. It would be interesting to address these questions more systematically in the future. This includes: 
\begin{itemize}
    
    \item We have used the known analytic behavior of the Toda system to obtain the time-averaged value of Krylov complexity when the temperature is very low, $\beta m\gg 1$. In this case the operator is confined to the first few sites of the ``Krylov chain'' and averaged $\bar K$ is exponentially small. It would be very interesting to develop analytic control over $\bar K$ when $\beta$ is only moderately large, $\beta m \sim 1$ to make a connection with  \cite{Chakraborty2024,Balasubramanian_2025}.  

    \item On a related note, a simple model for the 2pt function \eqref{toymeasure} provides a good qualitative description for Lanczos coefficients in field theory at various temperatures, making a connection with \cite{Vasli_2024,Anegawa:2024wov,Chattopadhyay:2024pdj,He:2024xjp,Aguilar-Gutierrez:2025kmw}. It would be interesting to bring this to a new level by making a connection with the quasi-normal mode expansion for the autocorrelation function developed in \cite{Dodelson:2024atp,Dodelson:2025rng}. 
    
    \item The equations describing $\beta$-dependence of $b_n$ can be used to numerically evaluate thermal 2pt function $C_\beta(t)$ for any values of $\beta$ starting from 
    the physical data for any other value $\beta=\beta_0$. In this paper we have illustrated this approach for a small spin chain with 5 sites, starting from infinite temperature $\beta_0=0$. It would be interesting to further develop this method to apply to larger system sizes, and potentially compete with other numerical approaches, such as direct diagonalization, matrix product states, quantum typicality and others \cite{PhysRevLett.93.040502,verstraete2004matrix,PhysRevLett.108.240401,elsayed2013regression,spin12,Richter}  to investigate finite temperature dynamics. 

    \item We have seen that the self-consistency of the equations \eqref{MTdynamics} driving $\beta$-dependence of $b_n$ ``knows'' a great deal about the underlying system. Thus for a 2d CFT consistency requires the spectrum to be degenerate. It would be interesting to generalize this approach, which we call ``Krylov bootstrap'', to other physical systems. 
    
     \item The Lanczos algorithm is a powerful numerical method to obtain the extreme eigenvalues \cite{cullum2002lanczos}. A remarkable feature, discussed in section  \ref{asymptoticsec} is that the Toda flow approaches the eigenvalues asymptotically when $\beta$ is large.   It is tempting to interpret this flow in terms of the low temperature dynamics, in the spirit of \cite{motta2020determining}  potentially leading to novel quantum algorithms to determine the spectrum of the energy gaps. 
     
\end{itemize}
We hope to address some of these questions in the near future.

\section*{Acknowledgments}
A.D.~acknowledges support by the NSF under  grant  2310426. DC is supported by the S\~ao Paulo Research
Foundation (FAPESP) through the grant 2024/13100-8.

\appendix

\section{Modified Toda Dynamics\label{appB}}
The procedure described in the main text yields the following expression for the coefficients $c_i$ in 
 terms of the eigenvalues of $T_\even$,
\be c_i=2(-1)^{i+1}{e_{n-i}(\{\lambda_j\})\over e_{n-1}(\{\lambda_j\})},\ee
where $e_i$ denotes the elementary homogeneous symmetric polynomial of order $i$.
Our goal below is to is to rewrite the expression ${\cal H}'=\sum_k c_k I^\even_k$ in a simpler form.

The determinant of $T_{\even}$ is  a function of the integrals of motion $\mathcal I_k^{\text{even}}$, 
\be \mathcal C\equiv {\rm det}(T_{\rm even}) = \frac{ e_{n-1}(\{\lambda_j\})}{2n}  \sum_{k=1}^n k\, c_k\, \mathcal I_k^\even.\ee
This quantity vanishes on-shell, due to $T$ having a zero eigenvalue.

Let us define the generating function
\be G(z)=\det(1+z\, T_\even)=e^{z \mathcal I^\even_1}e^{-z^2 \mathcal I^\even_2}e^{z^3 \mathcal I^\even_3}\cdots.\ee
We can write the elementary polynomials as follows
\be e_i={1\over i!}G^{(i)}(0).\ee
The determinant can be written as
\be \mathcal C = {1\over n!}G^{(n)}(0).\ee
Note that it vanishes on-shell $\mathcal C|_{os}=0$.
We then calculate
\be {\partial\over\partial \mathcal I_i}G(z)=(-1)^{i+1} z^i G(z)+O(z^{n+1}).\ee
From here follows
\be {\partial \mathcal C\over \partial \mathcal I_i}=(-1)^{i+1}{d^n\over dz^n} (z^i G(z)+O(z^{n+1}))|_{z=0}= n! e_{n-i}(-1)^{i+1},\ee
and we find for $c_i$,
\be c_i= { 2  \over {\partial \mathcal C \over \partial \mathcal I_1}}{\partial \mathcal C\over \partial \mathcal I_i}.\ee
The Hamiltonian can now be written as
\be \mathcal H'={2 \over {\partial \mathcal C\over \partial \mathcal I_1}\bigg|_{os}}\sum {\partial \mathcal C \over \partial \mathcal I_i}\bigg|_{os}\mathcal I_i .\ee
The same flow is generated by a slightly different function, since it has the same gradient as ${\mathcal H}'$ on-shell, i.e.~on the surface of the constraint
\be \mathcal H'={2\mathcal C\over \partial \mathcal C/\partial \mathcal I_1}.\ee

\section{Quantum Harmonic Oscillator\label{appOsc}}

Consider the quantum harmonic oscillator $H=\omega({1\over 2}+a^\dagger a)$ with initial operator $A_0=x$. The Krylov space is two-dimensional $A_0=x$, $A_1=-ip$, with Lanczos coefficient $b_0=\omega$.

The matrix $T_K$ as defined in (\ref{submatrix}) is
\be T_K=\omega \coth{\beta\omega\over 2} \begin{pmatrix}
    1 & 0\\
    0 & 1
\end{pmatrix}.\ee
This means that $B=0$ and the equation (\ref{Tdegen}) becomes
\be \dot T_K=-T_K^2+Y.\label{Tkdot}\ee
The matrix $Y$ can be calculated independently from
\be Y_{ij}={1\over 4}\langle \{H,x\},\{H,x\}\rangle \delta_{ij}=\omega^2 { \cosh(\beta\omega)\over \cosh(\beta\omega) -1}\delta_{ij}.\ee
It is now straightforward to confirm that equation (\ref{Tkdot}) is satisfied.

\section{$XY$ model}
\label{XY}

Consider the integrable XY model with periodic boundary conditions, described by the Hamiltonian:
\be H=\sum_{j=1}^N[(1+\gamma)S_j^xS_{j+1}^x+(1-\gamma)S^y_j S^y_{j+1}]-h\sum_{j=1}^N S_j^z.\label{XYham}\ee
We consider the limit where $N\to\infty$.

Define the following auto-correlation function at inverse temperature $\beta$:
\be C_\beta(t)=\langle S_0^z S_0^z(t)\rangle={\tr{e^{-\beta H/2}S_0^ze^{-\beta H/2} S_0^z(t)}\over \tr(e^{-\beta H})}.\ee

The Hamiltonian (\ref{XYham}) can be diagonalized by the Jordan-Wigner transformation. The  quasi-particle energies are
\be \epsilon_k=\sqrt{(\cos k-h)^2+\gamma^2\sin^2 k}.\ee
In addition, define
\be \lambda_k={1\over 2}\arctan{\gamma\sin k\over\cos k-h}.\ee
Note that in all formulas of this section, we make the choice $\arctan(x)\in (0,\pi]$.

The auto-correlation function is given by \cite{spin12}
\be C_\beta(t)=m_z^2+\left[{1\over 2\pi}\int_0^\pi dk~\cos(\epsilon_k t)\sech({1\over 2}\beta\epsilon_k)\right]^2+\left[{1\over 2\pi}\int_0^\pi dk~\cos(2\lambda_k)\sin(t\epsilon_k)\sech({1\over 2}\beta\epsilon_k)\right]^2,\ee
where
\be m_z\equiv \langle S_0^z\rangle_\beta=-{1\over 2\pi}\int_0^\pi dk~\cos(2 \lambda_k ) \tanh \frac{\beta \epsilon_k}{2}. \label{magdef}\ee

In order to calculate the moments, and subsequently Lnczos coefficients numerically, we first Taylor expand
\be {1\over 2\pi}\int_0^\pi dk~\cos(\epsilon_k t)\sech({1\over 2}\beta\epsilon_k)=\sum_{n=0}^\infty u_{n} t^{n}\implies u_{2n}={(-1)^n\over 2\pi} \int_0^\pi dk~ {\epsilon_k^{2n}\sech({1\over 2}\beta \epsilon_k)\over (2n)!},\ee
\be {1\over 2\pi}\int_0^\pi dk~\cos(2\lambda_k)\sin(t\epsilon_k)\sech({1\over 2}\beta\epsilon_k)=\sum_{n=0}^\infty v_{n} t^{n}\implies v_{2n+1}={(-1)^n\over 2\pi} \int_0^\pi dk~ {\epsilon_k^{2n}\sech({1\over 2}\beta \epsilon_k)\cos(2\lambda_k)\over (2n+1)!}.\ee
The moments are now given by
\be \mu_{2n}={1\over M_0}\left(m_z^2\delta_{n,0}+n!\sum_{i=0}^{n}(u_iu_{n-i}+v_i v_{n-i})\right),\ee
where $M_0$ is a constant chosen such that $\mu_0=1$.

\section{Asymptotic behavior of $b_o,b_e$ from the Dyck paths}
\label{Dyck}
In the  Dyck paths approach, the moments for large $n$ are given by the following path integral \cite{avdoshkin2022krylov}
\bea
\mu_{2n} = \int \mathcal{D} f(t) e^{n \int\limits_{0}^1 dt \left( W_{\text{eff}}(f', \epsilon(2nf)) + 2 \log b(2nf) \right)}, \label{mupath}
\eea
where the effective action
\bea
\nonumber
W_{\text{eff}}(f', \epsilon) = -\frac{f'}{2}\log \frac{\e^{{4}} f'^2+f' (f' + \sqrt{4 \e^2 + (\e^2-1)^2 f'^2})+\e^2 (2 + f' \sqrt{4 \e^2 + (\e^2-1)^2f'^2})}{2 \e^2 (1 - f')^2} +\\ {2\log(1+\e)}
+ \log \frac{(\e - 1)^2}{\e + \e^3 - \e\sqrt{4 \e^2 +(\e^2-1)^2 (f')^2}}.\qquad \qquad 
\label{Weff}
\eea
It is a function of 
\bea
\label{epsilon}
\epsilon(n)=b_{\text{even}}(n)/b_{\text{odd}}(n),\\
b(n)=\sqrt{b_{\text{even}}(n) b_{\text{odd}}(n)},
\label{bdef}
\eea
where it is assumed that  $b_n$ form two continuous branches for large $n$, $b_{2n} = b_{\text{even}}(2n)$ and $b_{2n+1} = b_{\text{odd}}(2n)$. Function $f(t)$ should satisfy  boundary conditions $f(0)=f(1)=0$ \cite{avdoshkin2020}.

With the additional assumption that  $b_{\text{even}}(n)$ and $b_{\text{odd}}(n)$ asymptote to constants, $b_o$ and $b_e$ respectively, we consider 
\bea
\epsilon=b_{e}/b_{o},\quad b=\sqrt{b_{e}b_o}
\eea
to be constant. Plugging this back into \eqref{Weff} and  \eqref{mupath} we find the solution for $f(t)=0$ and
\bea
\mu_{2n}=\left({(1+\e)^2\over \e}b^2\right)^n.
\eea
This matches leading exponential behavior $\mu_{2n}\approx \omega_{\text{max}}^{2n}$ associated with $\Phi(\omega)$ with a hard cutoff $\omega_{\text{max}}$. This is consistent with \eqref{asympt} for any $\Delta$, implying only that $b_e+b_o=\omega_{\text{max}}$.

\bibliographystyle{JHEP}
\bibliography{tempkrylov.bib}

\end{document}